\documentclass[12pt]{article}
\usepackage{amssymb}
\usepackage{amsfonts}
\usepackage{amsmath}
\usepackage{epsfig,bm,url,blkarray}
\usepackage[onehalfspacing]{setspace}
\usepackage{geometry}
\usepackage{lscape}
\usepackage{xcolor}
\usepackage{caption}
\usepackage[round]{natbib}
\usepackage{float} 
\usepackage{graphicx}

\renewcommand{\thesection}{\arabic{section}}

\usepackage{authblk}

\usepackage{xr}
\makeatletter
\newcommand*{\addFileDependency}[1]{
  \typeout{(#1)}
  \@addtofilelist{#1}
  \IfFileExists{#1}{}{\typeout{No file #1.}}
}
\makeatother

\newcommand*{\myexternaldocument}[1]{
    \externaldocument{#1}
    \addFileDependency{#1.tex}
    \addFileDependency{#1.aux}
}

\myexternaldocument{OnlineSupplement}

\begin{document}

\title{Enhancing Efficiency of Local Projections Estimation with Volatility Clustering in High-Frequency Data}
\author{Chew Lian Chua, David Gunawan and Sandy Suardi\thanks{Chua: School of Economics, University of Nottingham Ningbo, China. Gunawan: School of Mathematics and Applied Statistics, University of Wollongong, Australia. Suardi: School of Business, University of Wollongong, Australia.  }}
\maketitle

\begin{abstract}
This paper advances the local projections (LP) method by addressing its inefficiency in high-frequency economic and financial data with volatility clustering. We incorporate a generalized autoregressive conditional heteroskedasticity (GARCH) process to resolve serial correlation issues and extend the model with GARCH-X and GARCH-HAR structures. Monte Carlo simulations show that exploiting serial dependence in LP error structures improves efficiency across forecast horizons, remains robust to persistent volatility, and yields greater gains as sample size increases. Our findings contribute to refining LP estimation, enhancing its applicability in analyzing economic interventions and financial market dynamics.

Keywords: Local projection, GARCH, High frequency

JEL codes: C53, C22, C32
\end{abstract}

\section{Introduction}
The causal impact of interventions is central to applied economics. The local projections (LP) method, introduced by \cite{10.1257/0002828053828518}, has gained widespread use, typically applied with pre-identified structural shocks rather than identifying them during estimation. LP has been used to analyze US monetary policy's impact on global real exchange rates \citep{MA2024111891}, assess tax progressivity's effects on US income inequality since 1970 \citep{JALLES2024111715}, and explore crypto shocks' spillover into the US stock market \citep{MUSHOLOMBO2023111427}.

\cite{LI2024105722} show that LPs exhibit a bias-variance trade off in impulse response estimation, finding that the least-squares LP estimator has lower bias than the least-squares vector autoregressive (VAR) estimator but suffers from higher variance, reducing efficiency. This inefficiency arises because LP error terms across horizons are uncorrelated, unlike VAR, where impulse response variance at future horizon depends on error terms from previous horizons. To improve efficiency, much research has focused on refining LP estimators, including the use of HAC and HAR standard errors \citep{stock2018identification}. More recently,  \cite{https://doi.org/10.3982/ECTA18756} show that incorporating lagged variables in LP regressions eliminates the need for standard error corrections. 

This paper proposes LP models designed for high-frequency economic and financial data, which often exhibit volatility clustering. Our approach captures volatility clustering in the error term using a generalized autoregressive conditional heteroskedasticity (GARCH) process, addressing the serial correlation issue in LP models. We further extend the model by integrating local projection errors into the conditional covariance equation, incorporating exogenous terms (GARCH-X) and a heterogeneous structure (GARCH-HAR). By leveraging the serial dependence in LP error structures, Monte Carlo simulations demonstrate that our approach improves efficiency relative to standard LP across various forecast horizons, remains robust to persistent volatility, and yields greater gains as sample size increases. 

The rest of the article is organized as follows.
Section~\ref{sec:models}
discusses the proposed LP models. Section \ref{sec:simulation study} 
presents results comparing different LP models using simulated datasets. Section \ref{sec:conclusion}  concludes
with a discussion of our approach and results. This article has an online Appendix containing additional technical details and empirical results.


\section{Models\label{sec:models}} 
This section describes the proposed LP models and how the impulse
responses are generated from these models. We consider an $N\times1$
financial and economic variables $y_{t}=\left(y_{1t},...,y_{Nt}\right)^{\top}$
at time $t$. For ease of exposition, we assume $N=1$. The standard
LP model for $h$-step projection is given by 
\begin{equation}
y_{t+h}=c_{h}+\beta_{h}y_{t}+e_{h,t},\;e_{h,t}\sim N\left(0,\sigma^{2}\right),\label{eq:standardLPmodel}
\end{equation}
where $c_{h}$ is the intercept term, $\beta_{h}$ is the impulse
response at horizon $h$, and $e_{h,t}$ is the $h$-step projection
error term, which is assumed to follow a normal distribution with
zero mean and variance $\sigma^{2}$, and is correlated with the past
errors. Eq. \eqref{eq:standardLPmodel} shows that the Local Projection approach estimates a separate regression
for each time horizon $h$ to estimate impulse response. In general, the LP model yields higher impulse response variance than the VAR model \citep{LI2024105722} due to unmodeled serial correlation in LP error terms across different horizons.


In this paper, we extend the standard LP model in Eq. \eqref{eq:standardLPmodel}
by modeling the local projection errors at $h$-step projection to
follow a GARCH type model, 
\begin{eqnarray}
y_{t+h} & = & c_{h}+\beta_{h}y_{t}+e_{h,t},\;e_{h,t}\sim N\left(0,\sigma_{h,t}^{2}\right),\label{eq:LPGARCH1}\\
\sigma_{h,t}^{2} & = & \gamma_{h}+\alpha_{1,h}\sigma_{h,t-1}^{2}+\alpha_{2,h}e_{h,t-1}^{2}.\label{LPGARCH2}
\end{eqnarray}
The model in Eqs. \eqref{eq:LPGARCH1} and \eqref{LPGARCH2} is the LP-GARCH model.


The LP-GARCH model in Eqs. \eqref{eq:LPGARCH1} and \eqref{LPGARCH2}
does not account for serial correlation in the h-step projection error term $e_{h,t}$.
Next, we introduce the LP-GARCHX model, where the conditional variance $\sigma_{h,t}^{2}$
depends on both the squared projection errors from the $h-1$ step, $e_{h-1,t}^{2}$, and the standard GARCH components, $\sigma_{h,t-1}^{2}$ and $e_{h,t-1}^{2}$.
The LP-GARCHX is given by
\begin{eqnarray}
y_{t+h} & = & c_{h}+\beta_{h}y_{t}+e_{h,t},\;e_{h,t}\sim N\left(0,\sigma_{h,t}^{2}\right),\label{eq:LPGARCHX-1}\\
\sigma_{h,t}^{2} & = & \gamma_{h}+\alpha_{1,h}\sigma_{h,t-1}^{2}+\alpha_{2,h}e_{h,t-1}^{2}+\alpha_{3,h}e_{h-1,t}^{2}.\label{eq:LPGARCHX-2}
\end{eqnarray}
Note that when $h=1$, the parameter $\alpha_{3,h}=0$. As $\sigma_{h,t}^{2}$
depends on $e_{h-1,t}$ from $h=2$ onwards, 
we need a set of estimates of $e_{h-1,t}$ to estimate $\sigma_{h,t}^{2}$. We can proceed with a
recursive strategy for estimating LP equations starting from $h=1$
to the end of the forecast horizon. The steps are:
\begin{enumerate}
\item At $h=1$, estimate the LP-GARCH equation 
\begin{eqnarray*}
y_{t+h} & = & c_{h}+\beta_{h}y_{t}+e_{h,t},\\
\sigma_{h,t}^{2} & = & \gamma_{h}+\alpha_{1,h}\sigma_{h,t-1}^{2}+\alpha_{2,h}e_{h,t-1}^{2},
\end{eqnarray*}
\item Obtain the estimates of $e_{h,t}$,
\begin{equation}
\widehat{e}_{h,t}=y_{t+h}-\widehat{c}_{h}-\widehat{\beta}_{h}y_{t}.
\end{equation}
\item Increment $h$ by $1$, and estimate the following LP equation
\begin{eqnarray*}
y_{t+h} & = & c_{h}+\beta_{h}y_{t}+e_{h,t},\\
\sigma_{h,t}^{2} & = & \gamma_{h}+\alpha_{1,h}\sigma_{h,t-1}^{2}+\alpha_{2,h}e_{h,t-1}^{2}+\alpha_{3,h}\widehat{e}_{h-1,t}^{2},
\end{eqnarray*}
\item Repeat Steps 2 and 3 until the end of the forecast horizon. 
\end{enumerate}

The final model, LP-GARCH-HAR, combines the GARCH framework with Heterogeneous Autoregressive (HAR) model of \citet{corsi2009simple} to capture long-memory effects commonly observed in high-frequency financial time series.
The LP-GARCH-HAR model is given by 
\begin{eqnarray}
y_{t+h} & = & c_{h}+\beta_{h}y_{t}+e_{h,t},\;e_{h,t}\sim N\left(0,\sigma_{h,t}^{2}\right),\label{eq:LPGARCH-HAR-1}\\
\sigma_{h,t}^{2} & = & \gamma_{h}+\alpha_{1,h}\sigma_{h,t-1}^{2}+\alpha_{2,h}e_{h,t-1}^{2}+\alpha_{3,h}e_{h-1,t}^{2}+\alpha_{4,h}\widetilde{e}_{h-1,t}^{2}+1_{\left(h-1\right)>5}\alpha_{5,h}\overline{e}_{h-5,t}^{2},\label{eq:LPGARCH-HAR-2}
\end{eqnarray}
where $1_{\left(h-1\right)>5}$ is an indicator function that equals
$1$ when $h-1>5$, and $0$ otherwise, 
\begin{equation}
\widetilde{e}_{h-1,t}^{2}=\frac{1}{h-1}\sum_{i=1}^{h-1}e_{i,t}^{2},
\end{equation}
and 
\begin{equation}
\overline{e}_{h-5,t}^{2}=\frac{1}{5}\sum_{i=1}^{5}e_{i,t}^{2}.
\end{equation}
The LP models are estimated using the Maximum Likelihood (ML) method, with optimization performed in Matlab. Section \ref{sec:simulation study} evaluates the efficiency of the proposed LP models through a Monte Carlo study.


\section{Simulation Study\label{sec:simulation study}}
This section evaluates the efficiency of different LP models in the Monte Carlo study. The Monte Carlo design is described in Section \ref{sec:MCdesign}. Section \ref{sec:Results} discusses the results. 

\subsection{Monte Carlo Design\label{sec:MCdesign}}
The data generating process (DGP) follows a first-order autoregressive (AR(1)) model with a time-varying conditional variance governed by a generalized autoregressive conditional heteroskedasticity (GARCH(1,1)) process:
\begin{eqnarray}
y_{t} & = & \beta_{0}+\beta_{1}y_{t-1}+\epsilon_{t},\;t=1,...,T,\\
\epsilon_{t} & \sim & N\left(0,\sigma_{t}^{2}\right),\;t=1,...,T,\\
\sigma_{t}^{2} & = & \gamma+\alpha_{1}\sigma_{t-1}^{2}+\alpha_{2}\epsilon_{t-1}^{2}, \;t=2,...,T.
\end{eqnarray}
with the initial conditional variance 
\begin{equation}
\sigma_{1}^{2}=\frac{\gamma}{1-\alpha_{1}-\alpha_{2}}.
\end{equation}
We set the true parameter values to $\beta_0=0$, $\gamma=1$, $\alpha_1=0.5$, $\alpha_2=0.3,0.4,0.48$, and $\beta_{1} \in \{0.6, 0.8, 0.9, 0.95 \}$. For each true model, we generate $R=500$ datasets with lengths $T = 500, 1000, 2000, 5000$. The standard errors, derived from the specified true model, serve as benchmarks for evaluating four LP variants: (1) standard LP, (2) LP-GARCH, (3) LP-GARCHX, and (4) LP-GARCH-HAR (see Section \ref{sec:models}). We compare the impulse response standard errors from these LP models against those from the true models to assess their relative accuracy. The standard errors for the estimates of $\beta_{h}$ for $h=1,...,24$ obtained from different LP models and the true model are calculated as the standard deviation of the estimates $\beta^{(r)}_{h}$ for $r=1,...,R$ and $h=1,...,24$. For robustness in the standard errors, we examine different levels of persistence in the mean process by varying $\beta_1$ and in the GARCH process by adjusting $\alpha_2$.

\subsection{Results\label{sec:Results}}
\begin{figure}[H]
\caption{The standard errors of the estimated impulse responses for $h=1,...,24$
steps ahead for four LP models and the AR(1)-GARCH(1,1), with 
$T=500,1000,2000,5000$, $\beta_{1}=0.95$, $\alpha_1=0.5$ and $ \alpha_2=0.4$.\label{fig:Thestderrors095}}

\centering{}\includegraphics[width=15cm,height=8cm]{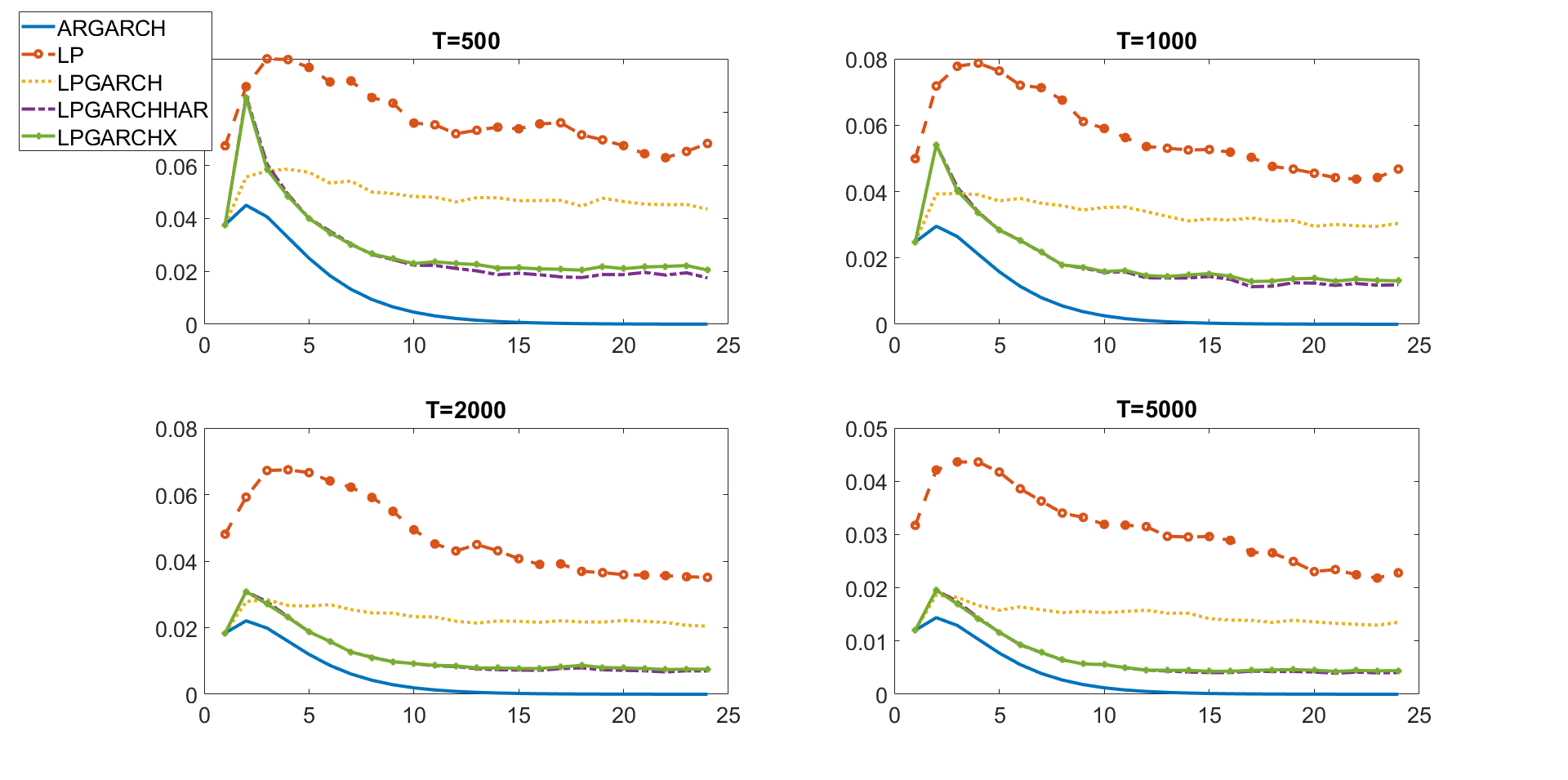}
\end{figure}

Figure \ref{fig:Thestderrors095} reports the standard errors of impulse responses for $h=1,…,24$  steps ahead across four LP models and the AR(1)-GARCH(1,1) model, considering different sample sizes with $\beta_1=0.95$, $\alpha_1=0.5$, and $\alpha_2=0.4$. For brevity, results for $\beta_1=0.6, 0.8, 0.9$ are provided in the online Appendix (Figures \ref{fig:Thestderrors06} to \ref{fig:Thestderrors09} in Section \ref{sec:Additionalresultsalpha2_0.4}). Overall, the LP-GARCH model yields smaller standard errors than the standard LP model. Additionally, the LP-GARCHX and LP-GARCH-HAR models exhibit similar standard errors, consistently outperforming the LP-GARCH model in terms of efficiency. 

\begin{figure}[H]
\caption{The differences in standard errors of the estimated impulse responses
for $h=1,...,24$ steps ahead for four LP models and the AR(1)-GARCH(1,1), with
 $T=500,1000,2000,5000$, $\beta_{1}=0.95$, $\alpha_1=0.5$ and $ \alpha_2=0.4$.\label{fig:diffstderror095}}

\centering{}\includegraphics[width=15cm,height=8cm]{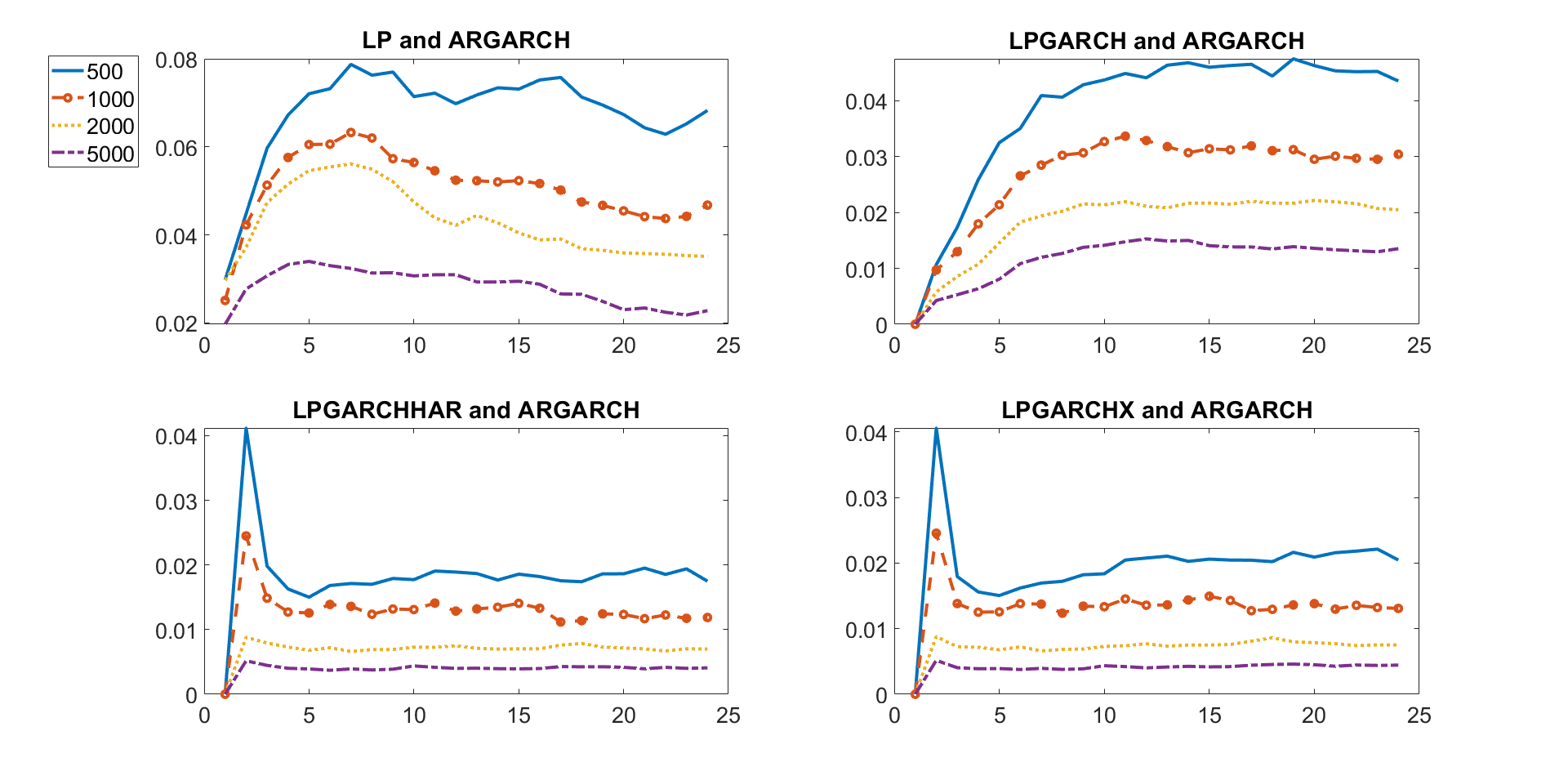}
\end{figure}

To assess the relative accuracy of standard errors across the four LP models compared to the AR(1)-GARCH(1,1) data-generating process (DGP), Figure \ref{fig:diffstderror095} plots the differences in standard errors of the estimated impulse responses. The gap between the LP models and the true model narrows as sample size increases, indicating that LP estimator efficiency improves with larger samples. Corresponding results for $\beta_1=0.6,0.8,0.9$ are provided in the online Appendix (Figures \ref{fig:diffstderror06} to \ref{fig:diffstderror09} in Section \ref{sec:Additionalresultsalpha2_0.4}).


Table \ref{tab:Root-Mean-Square LP} presents the mean standard errors relative to the true model (AR(1)-GARCH(1,1)) across the 24-step forecast horizon for the four LP models. It shows that mean standard errors decrease as the sample size increases, and that the proposed LP-GARCH, LP-GARCHX, and LP-GARCH-HAR models consistently produce smaller standard errors compared to the standard LP model. This suggests that our proposed methods enhance the efficiency of LP estimators. 

Sections \ref{sec:Additionalresultsalpha2_0.3} and \ref{sec:Additionalresultsalpha2_0.48} of the online Appendix present results for $\alpha_2=0.3$ and $\alpha_2=0.48$, respectively. 
Table \ref{tab:Root-Mean-Square LP}, along with Tables \ref{tab:Root-Mean-Square LP (alpha2_03)} and \ref{tab:Root-Mean-Square LP (alpha2_048)} in the online Appendix, shows that for standard LP models, mean standard errors relative to the true model are larger when $\alpha_2=0.48$ (i.e., a more persistent volatility process) and slightly smaller or unchanged when $\alpha_2=0.3$ (i.e., a less persistent volatility process). In contrast, for LP-GARCH, LP-GARCHX, and LP-GARCH-HAR models, the mean standard errors remain stable across different volatility persistence governed by the various $\alpha_2$ values.

In summary, the simulation study suggests that: (1) The proposed LP models, LP-GARCHX and LP-GARCH-HAR exhibit smaller standard errors than the LP-GARCH and standard LP models; (2) The gap measuring the differences in standard errors of the estimated impulse responses between the LP models and the true model decreases as sample size increases, indicating that LP estimator efficiency improves with larger samples; (3) For the standard LP model, the mean standard errors relative to the true model tend to be larger when $\alpha_2=0.48$ and smaller when $\alpha_2=0.3$. In contrast, for the LP-GARCH, LP-GARCHX, and LP-GARCH-HAR models, the mean standard errors relative to the true model remain consistent regardless of persistence in the GARCH process (i.e., variations in $\alpha_2$).



\begin{table}[H]
\caption{Mean standard errors of various Local Projection models relative to the AR(1)-GARCH(1,1), averaged over a 24-step forecast horizon. The GARCH data-generating process (DGP) is defined by $\gamma=1$, $\alpha_1=0.5$, $\alpha_2=0.4$.\label{tab:Root-Mean-Square LP} }
\centering{}%
\begin{tabular}{cccccc}
\hline 
$\beta_{1}$ & T & LP & LP-GARCH & LP-GARCH-HAR & LP-GARCHX\tabularnewline
\hline 
0.6 & 500 & $0.0495$ & $0.0478$ & $0.0274$ & $0.0275$\tabularnewline
 & 1000 & $0.0393$ & $0.0350$ & $0.0196$ & $0.0202$\tabularnewline
 & 2000 & $0.0297$ & $0.0249$ & $0.0129$ & $0.0134$\tabularnewline
 & 5000 & $0.0268$ & $0.0162$ & $0.0090$ & $0.0094$\tabularnewline
\hline 
0.8 & 500 & $0.0575$ & $0.0422$ & $0.0250$ & $0.0256$\tabularnewline
 & 1000 & $0.0477$ & $0.0280$ & $0.0173$ & $0.0179$\tabularnewline
 & 2000 & $0.0421$ & $0.0218$ & $0.0122$ & $0.0128$\tabularnewline
 & 5000 & $0.0313$ & $0.0133$ & $0.0066$ & $0.0070$\tabularnewline
\hline 
0.9 & 500 & $0.0726$ & $0.0434$ & $0.0243$ & $0.0262$\tabularnewline
 & 1000 & $0.0564$ & $0.0287$ & $0.0150$ & $0.0158$\tabularnewline
 & 2000 & $0.0452$ & $0.0209$ & $0.0102$ & $0.0106$\tabularnewline
 & 5000 & $0.0335$ & $0.0129$ & $0.0058$ & $0.0061$\tabularnewline
\hline 
0.95 & 500 & $0.0679$ & $0.0386$ & $0.0182$ & $0.0195$\tabularnewline
 & 1000 & $0.0509$ & $0.0269$ & $0.0128$ & $0.0134$\tabularnewline
 & 2000 & $0.0429$ & $0.0184$ & $0.0069$ & $0.0072$\tabularnewline
 & 5000 & $0.0281$ & $0.0118$ & $0.0039$ & $0.0041$\tabularnewline
\hline 
\end{tabular}
\end{table}

\section{Conclusion\label{sec:conclusion}}
In conclusion, this paper enhances the LP method by addressing its inefficiency in high-frequency economic and financial data exhibiting volatility clustering. By incorporating a GARCH process and extending the model with GARCH-X and GARCH-HAR structures, we capture the serial correlation in the local projection errors. Monte Carlo simulations demonstrate that exploiting serial dependence in LP error structures improves efficiency across forecast horizons, remains robust to persistent volatility, and yields greater gains as sample size increases. These findings refine LP estimation, broadening its applicability in analyzing high-frequency economic and financial data.

\pagebreak

\appendix

\renewcommand{\thealgorithm}{S\arabic{algorithm}}
\renewcommand{\theequation}{S\arabic{equation}}
\renewcommand{\thesection}{S\arabic{section}}
\renewcommand{\thepage}{S\arabic{page}}
\renewcommand{\thetable}{S\arabic{table}}
\renewcommand{\thefigure}{S\arabic{section}.\arabic{figure}}
\renewcommand{\thelemma}{S\arabic{lemma}}
\setcounter{page}{1}
\setcounter{section}{0}
\setcounter{equation}{0}
\setcounter{table}{0}
\setcounter{figure}{0}

\begin{center}
    \Large \textbf{Online Appendix: Enhancing Efficiency of Local Projections Estimation with Volatility Clustering in High-Frequency Data}
\end{center}

In the online Appendix, we present the simulation results for different persistence in the mean process in Section \ref{sec:Additionalresultsalpha2_0.4}, and for varying persistence in the volatility process in Sections \ref{sec:Additionalresultsalpha2_0.3} and \ref{sec:Additionalresultsalpha2_0.48}. All other notations are consistent with those defined in the main paper.


\section{Additional Figures and Tables for $\alpha_2=0.4$\label{sec:Additionalresultsalpha2_0.4}}

\begin{figure}[H]
\caption{The standard errors of the estimated impulse responses for $h=1,...,24$
steps ahead for four LP models and the AR(1)-GARCH(1,1), with
$T=500,1000,2000,5000$, $\beta_{1}=0.6$, $\alpha_1=0.5$ and $\alpha_2=0.4$.\label{fig:Thestderrors06}}
\centering{}\includegraphics[width=15cm,height=8cm]{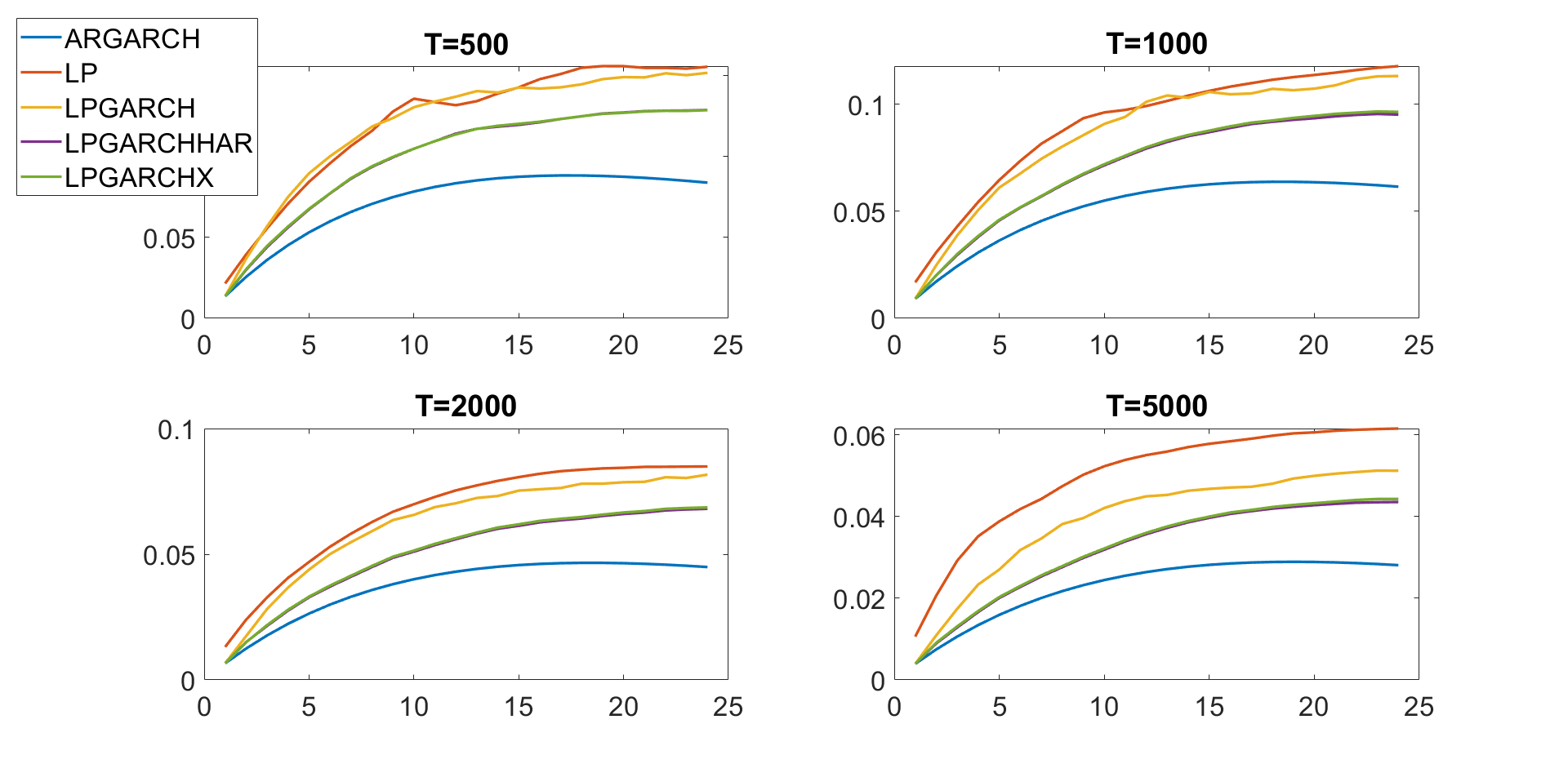}
\end{figure}

\begin{figure}[H]
\caption{The standard errors of the estimated impulse responses for $h=1,...,24$
steps ahead for four LP models and the AR(1)-GARCH(1,1), with
$T=500,1000,2000,5000$, $\beta_{1}=0.8$, $\alpha_1=0.5$ and $ \alpha_2=0.4$.\label{fig:Thestderrors08}}

\centering{}\includegraphics[width=15cm,height=8cm]{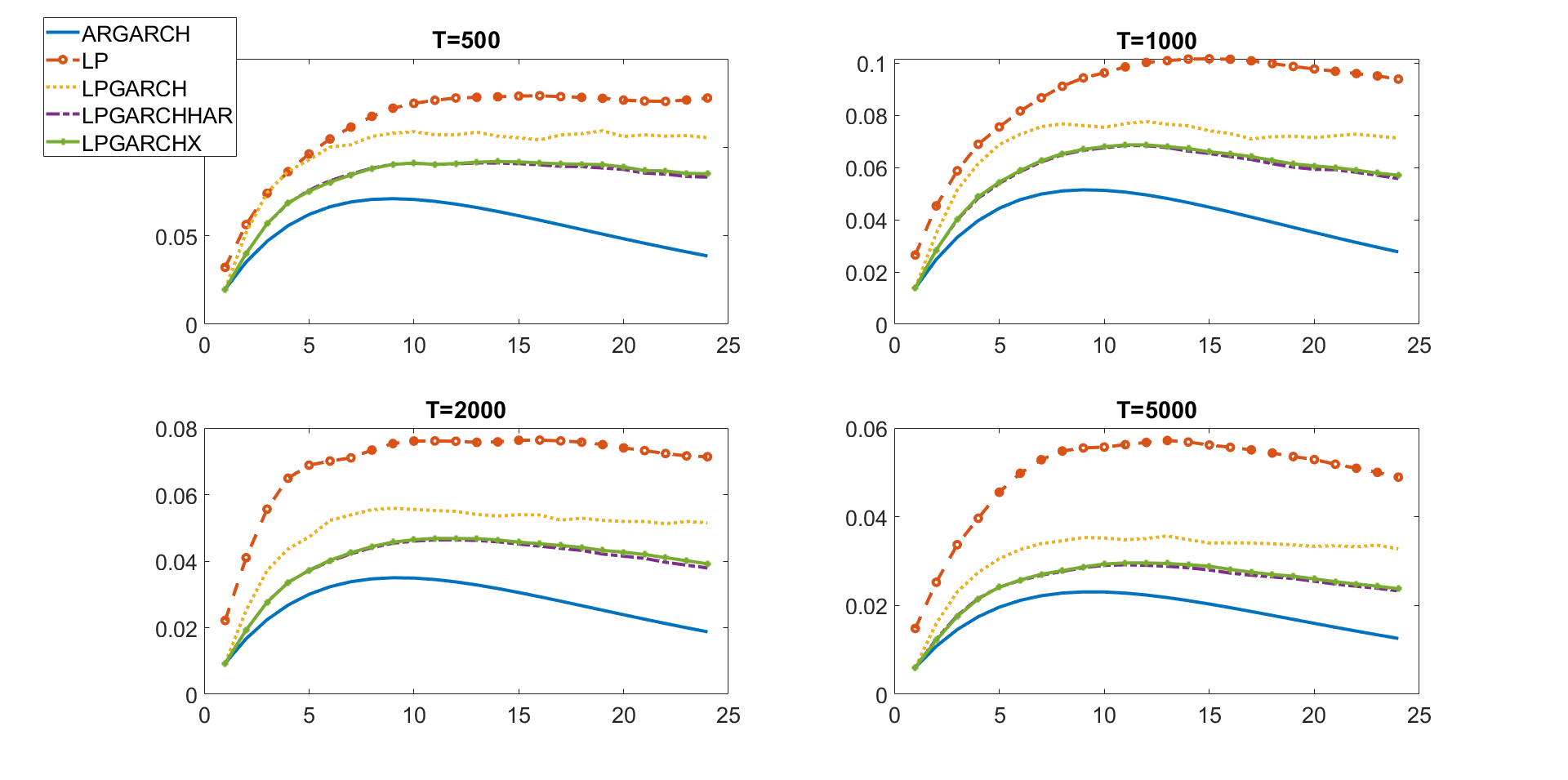}
\end{figure}

\begin{figure}[H]
\caption{The standard errors of the estimated impulse responses for $h=1,...,24$
steps ahead for four LP models and the AR(1)-GARCH(1,1), with
$T=500,1000,2000,5000$, $\beta_{1}=0.9$, $\alpha_1=0.5$ and $ \alpha_2=0.4$.\label{fig:Thestderrors09}}

\centering{}\includegraphics[width=15cm,height=8cm]{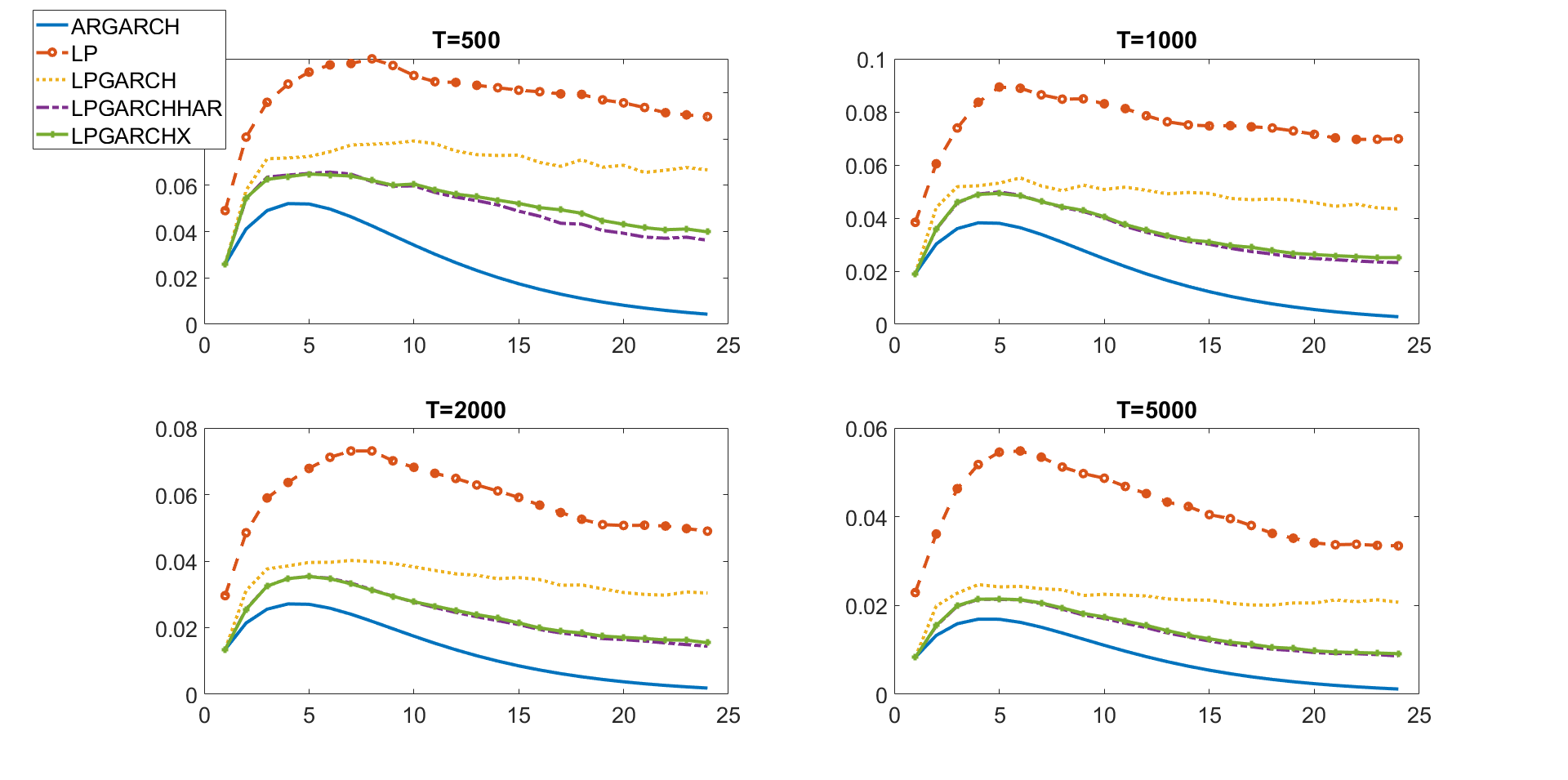}
\end{figure}

\begin{figure}[H]
\caption{The differences in standard errors of the estimated impulse responses
for $h=1,...,24$ steps ahead for four LP models and the AR(1)-GARCH(1,1),
with $T=500,1000,2000,5000$, $\beta_{1}=0.6$, $\alpha_1=0.5$ and $ \alpha_2=0.4$.\label{fig:diffstderror06}}

\centering{}\includegraphics[width=15cm,height=8cm]{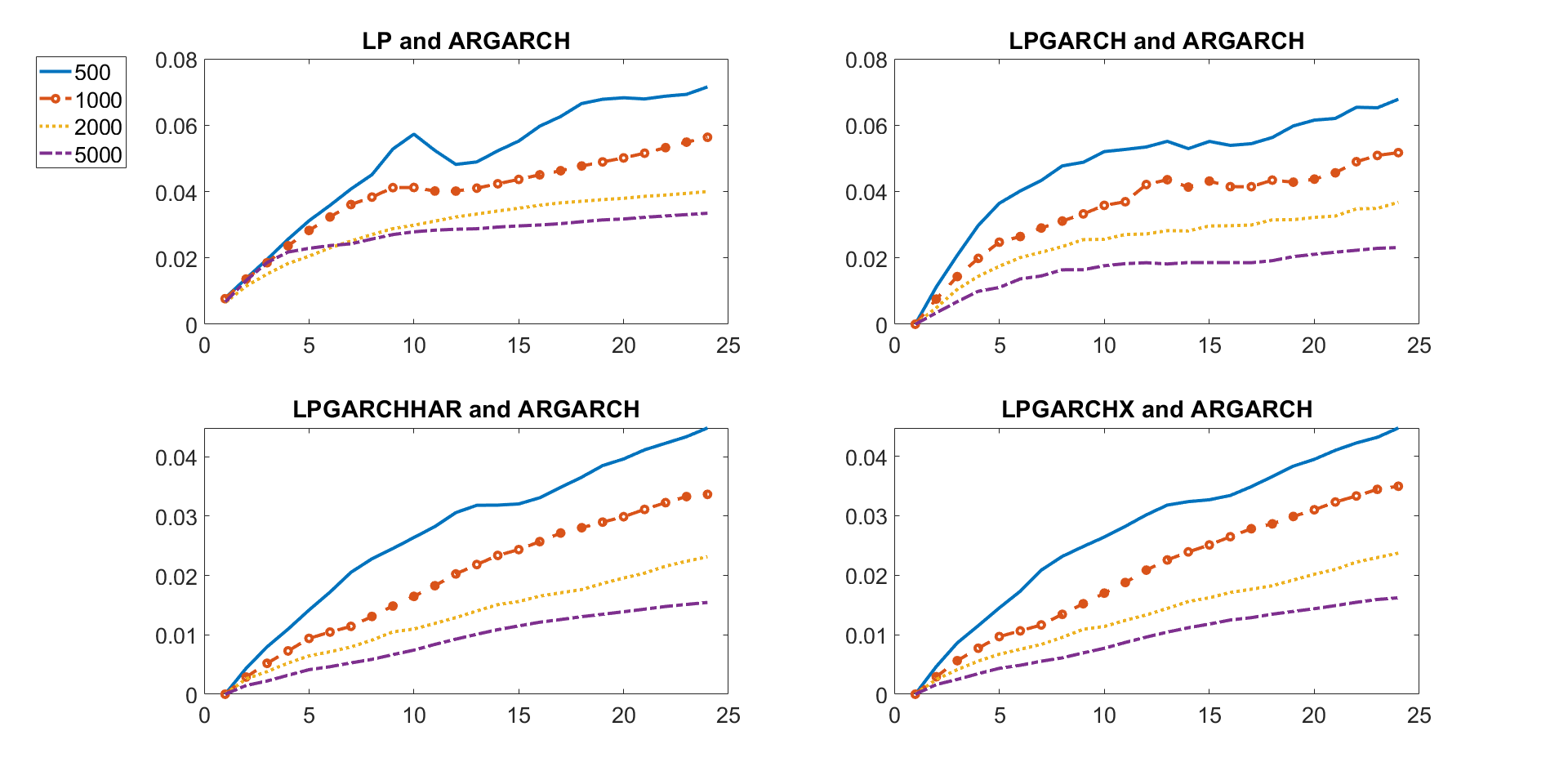}
\end{figure}

\begin{figure}[H]
\caption{The differences in standard errors of the estimated impulse responses
for $h=1,...,24$ steps ahead for four LP models and the AR(1)-GARCH(1,1), with $T=500,1000,2000,5000$, $\beta_{1}=0.8$, $\alpha_1=0.5$ and $\alpha_2=0.4$.\label{fig:diffstderror08}}

\centering{}\includegraphics[width=15cm,height=8cm]{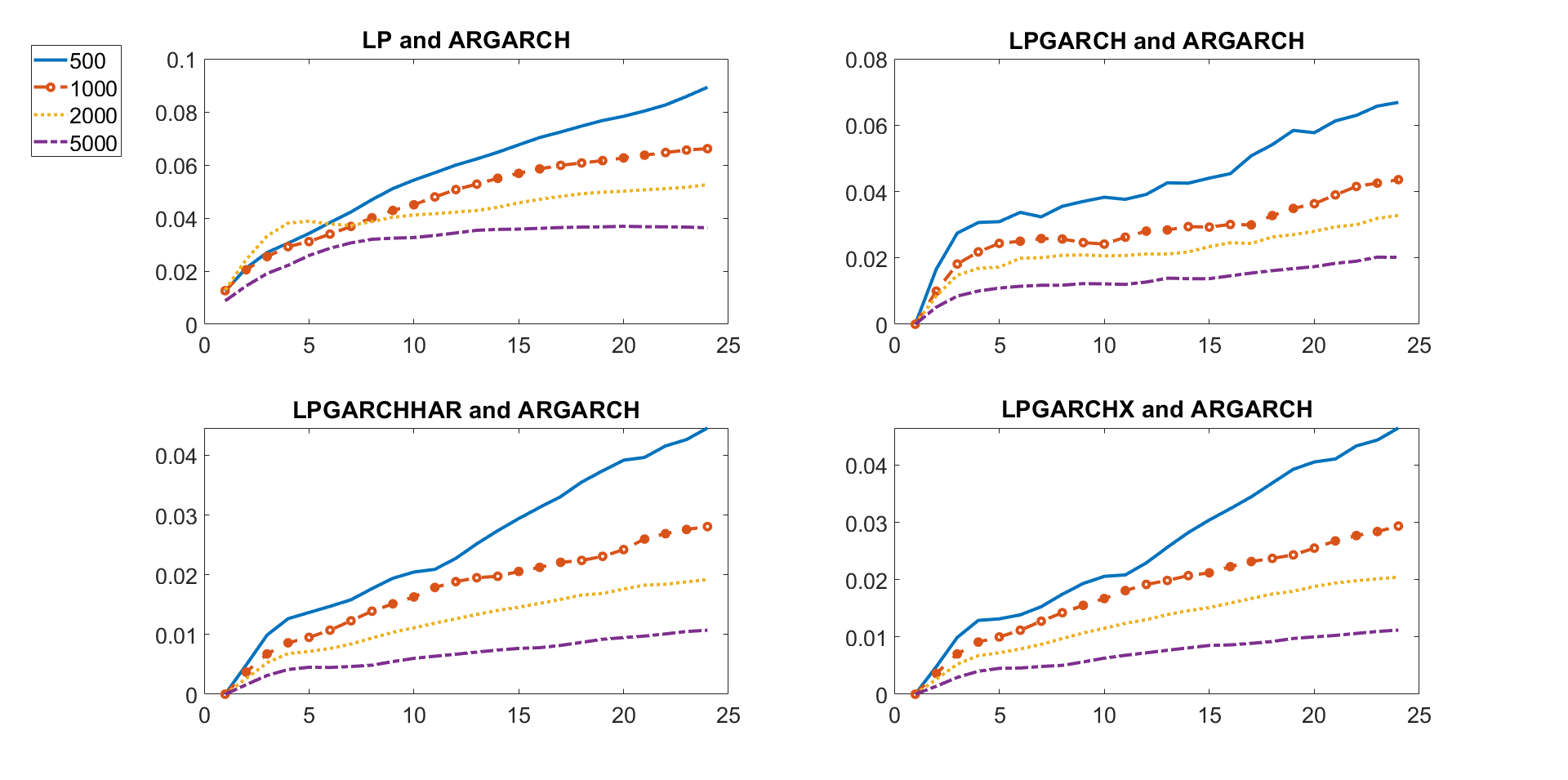}
\end{figure}

\begin{figure}[H]
\caption{The differences in standard errors of the estimated impulse responses
for $h=1,...,24$ steps ahead for four LP models and the AR(1)-GARCH(1,1), with
 $T=500,1000,2000,5000$, $\beta_{1}=0.9$, $\alpha_1=0.5$ and $ \alpha_2=0.4$.\label{fig:diffstderror09}}

\centering{}\includegraphics[width=15cm,height=8cm]{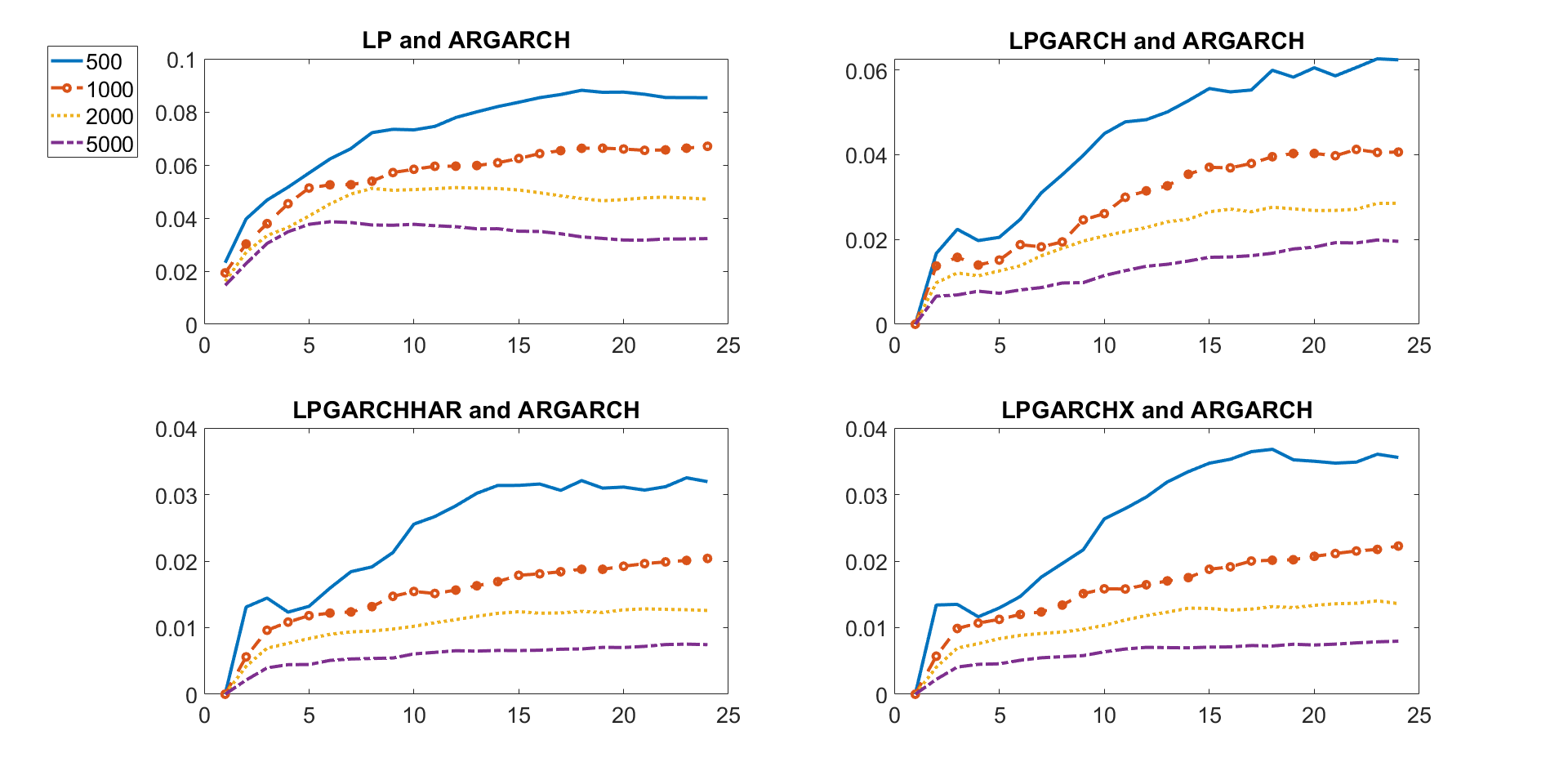}
\end{figure}

\setcounter{figure}{0}

\section{Additional Figures and Tables for $\alpha_2=0.3$ \label{sec:Additionalresultsalpha2_0.3}}

\begin{table}[H]
\caption{Mean standard errors of various Local Projection models relative to the AR(1)-GARCH(1,1), averaged over a 24-step forecast horizon. The GARCH data-generating process (DGP) is defined by $\gamma=1$, $\alpha_1=0.5$, $\alpha_2=0.3$.\label{tab:Root-Mean-Square LP (alpha2_03)}}



\centering{}%
\begin{tabular}{cccccc}
\hline 
$\beta_{1}$ & T & LP & LP-GARCH & LP-GARCH-HAR & LP-GARCHX\tabularnewline
\hline 
0.6 & 500 & $0.0382$ & $0.0473$ & $0.0267$ & $0.0268 $\tabularnewline
 & 1000 & $0.0256$ & $0.0334$ & $0.0183$ & $0.0188$\tabularnewline
 & 2000 & $0.0173$ & $0.0231$ & $0.0119$ & $0.0122$\tabularnewline
 & 5000 & $0.0130$ & $0.0155$ & $0.0085$ & $0.0087$\tabularnewline
\hline 
0.8 & 500 & $0.0478$ & $0.0438$ & $0.0252$ & $0.0259 $\tabularnewline
 & 1000 & $0.0365$ & $0.0290$ & $0.0170$ & $0.0174$\tabularnewline
 & 2000 & $0.0271$ & $0.0221$ & $0.0120$ & $0.0124$\tabularnewline
 & 5000 & $0.0174$ & $0.0139$ & $0.0064$ & $0.0066$\tabularnewline
\hline 
0.9 & 500 & $0.0627$ & $0.0473$ & $0.0251$ & $0.0264 $\tabularnewline
 & 1000 & $0.0454$ & $0.0317$ & $0.0156$ & $0.0161$\tabularnewline
 & 2000 & $0.0335$ & $0.0233$ & $0.0103$ & $0.0106$\tabularnewline
 & 5000 & $0.0213$ & $0.0143$ & $0.0056$ & $0.0057$\tabularnewline
\hline 
0.95 & 500 & $0.0528$ & $0.0429$ & $0.0187$ & $0.0199 $\tabularnewline
 & 1000 & $0.0421$ & $0.0300$ & $0.0119$ & $0.0125$\tabularnewline
 & 2000 & $0.0309$ & $0.0207$ & $0.0070$ & $0.0072$\tabularnewline
 & 5000 & $0.0193$ & $0.0133$ & $0.0039$ & $0.0040$\tabularnewline
\hline 
\end{tabular}
\end{table}

\begin{figure}[H]
\caption{The standard errors of the estimated impulse responses for $h=1,...,24$
steps ahead for four LP models and the AR(1)-GARCH(1,1), with
$T=500,1000,2000,5000$, $\beta_{1}=0.6$, $\alpha_1=0.5$ and $ \alpha_2=0.3$.\label{fig:Thestderrors06_alpha03}}
\centering{}\includegraphics[width=15cm,height=8cm]{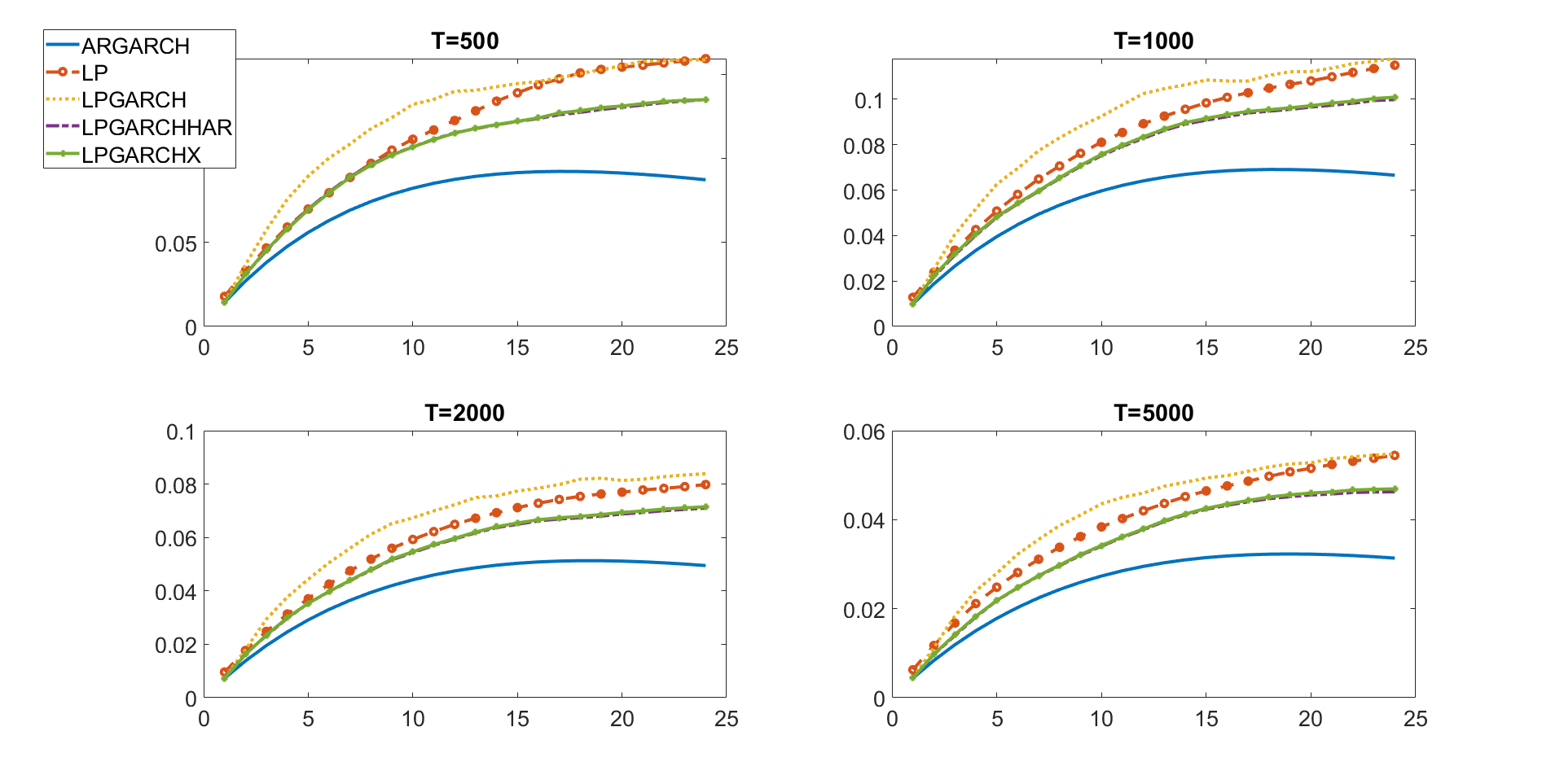}
\end{figure}

\begin{figure}[H]
\caption{The standard errors of the estimated impulse responses for $h=1,...,24$
steps ahead for four LP models and the AR(1)-GARCH(1,1), with
$T=500,1000,2000,5000$, $\beta_{1}=0.8$, $\alpha_1=0.5$ and $ \alpha_2=0.3$.\label{fig:Thestderrors08_alpha03}}

\centering{}\includegraphics[width=15cm,height=8cm]{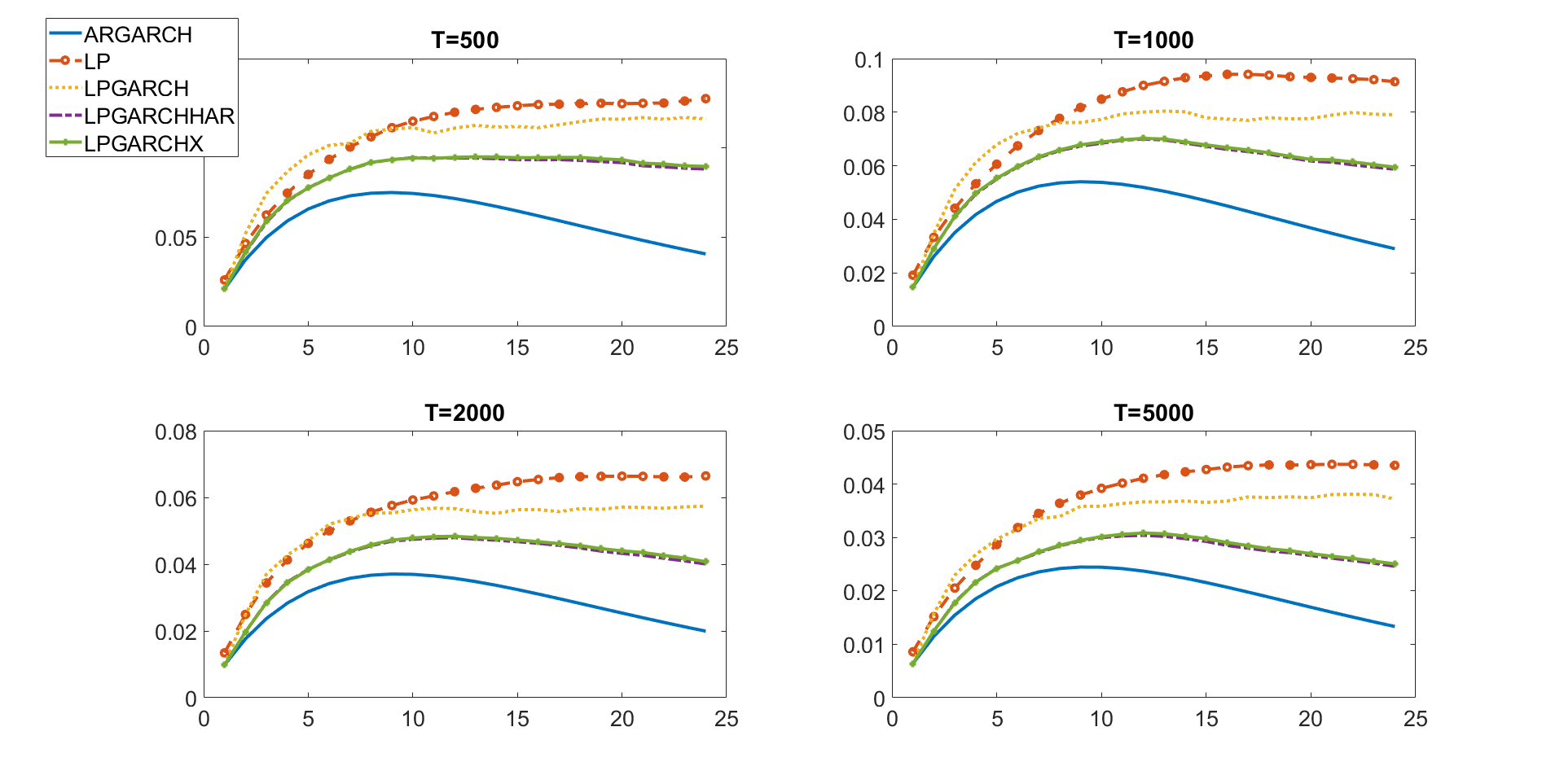}
\end{figure}

\begin{figure}[H]
\caption{The standard errors of the estimated impulse responses for $h=1,...,24$
steps ahead for four LP models and the AR(1)-GARCH(1,1), with
$T=500,1000,2000,5000$, $\beta_{1}=0.9$, $\alpha_1=0.5$ and $ \alpha_2=0.3$.\label{fig:Thestderrors09_alpha03}}

\centering{}\includegraphics[width=15cm,height=8cm]{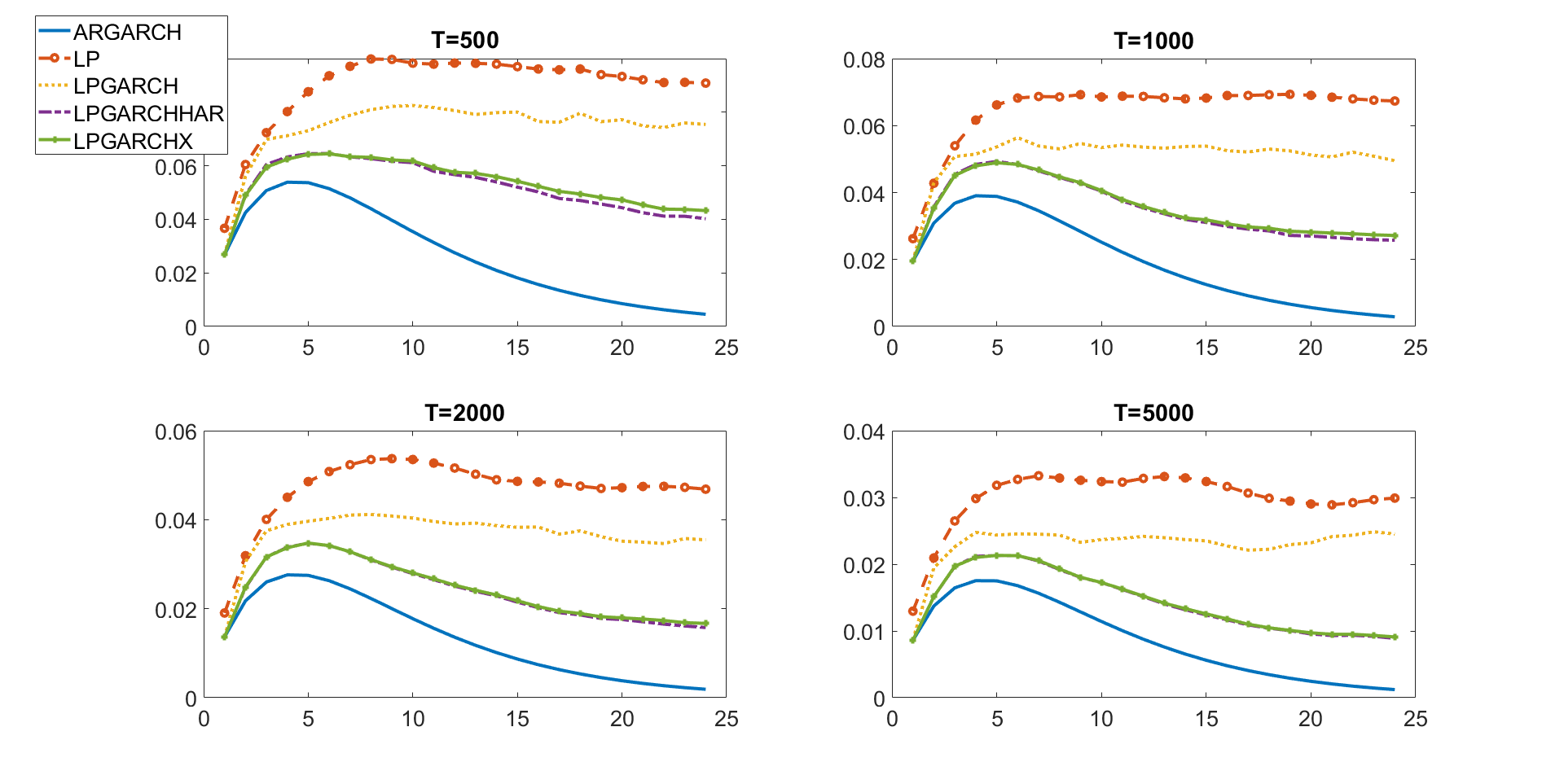}
\end{figure}

\begin{figure}[H]
\caption{The standard errors of the estimated impulse responses for $h=1,...,24$
steps ahead for four LP models and the AR(1)-GARCH(1,1), with
$T=500,1000,2000,5000$, $\beta_{1}=0.95$, $\alpha_1=0.5$ and $ \alpha_2=0.3$.\label{fig:Thestderrors095_alpha03}}

\centering{}\includegraphics[width=15cm,height=8cm]{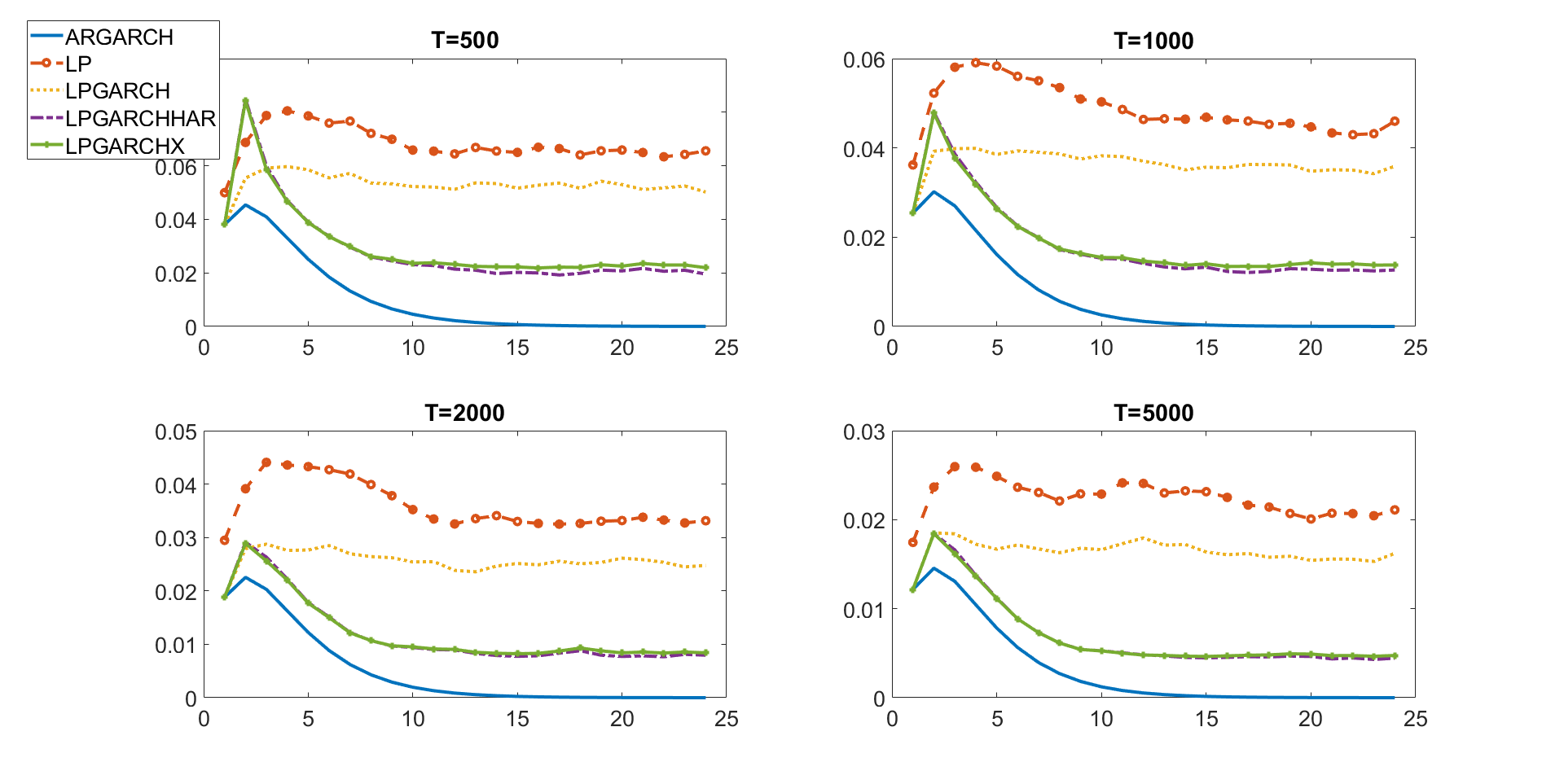}
\end{figure}

\begin{figure}[H]
\caption{The differences in standard errors of the estimated impulse responses
for $h=1,...,24$ steps ahead for four LP models and the AR(1)-GARCH(1,1), with
 $T=500,1000,2000,5000$, $\beta_{1}=0.6$, $\alpha_1=0.5$ and $ \alpha_2=0.3$.\label{fig:diffstderror06_alpha03}}

\centering{}\includegraphics[width=15cm,height=8cm]{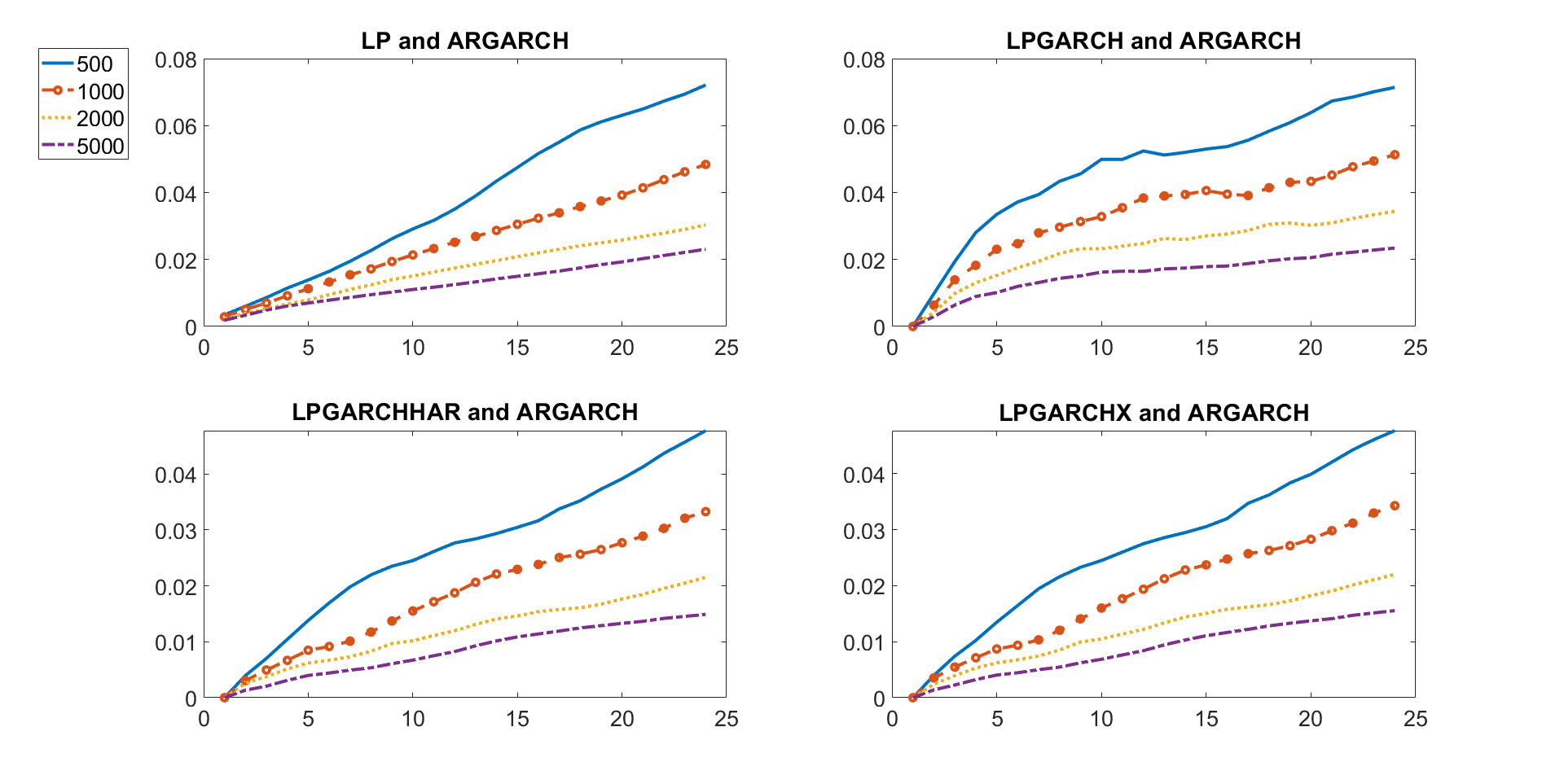}
\end{figure}

\begin{figure}[H]
\caption{The differences in standard errors of the estimated impulse responses
for $h=1,...,24$ steps ahead for four LP models and the AR(1)-GARCH(1,1), with
 $T=500,1000,2000,5000$, $\beta_{1}=0.8$, $\alpha_1=0.5$ and $ \alpha_2=0.3$.\label{fig:diffstderror08_alpha03}}

\centering{}\includegraphics[width=15cm,height=8cm]{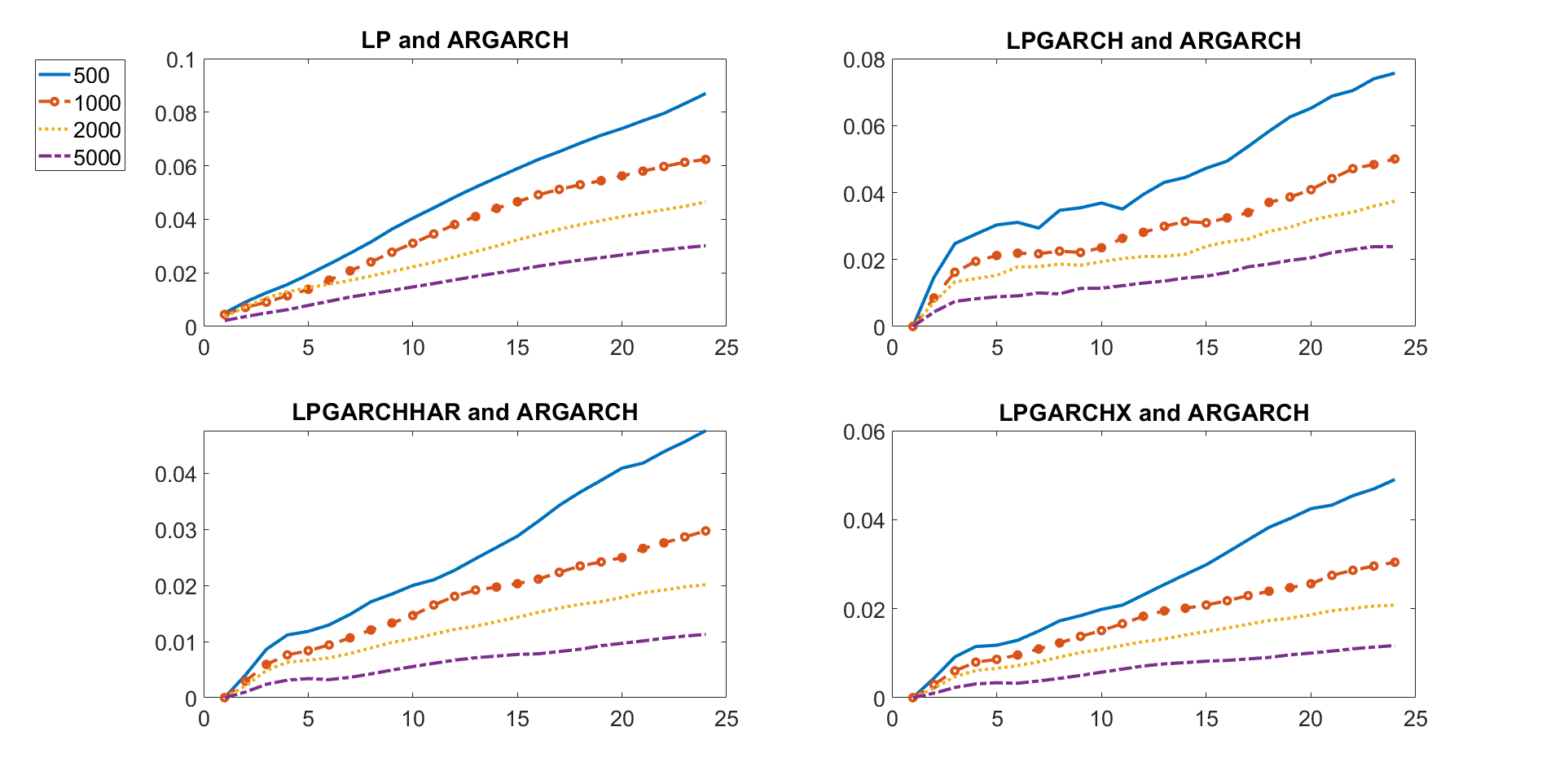}
\end{figure}

\begin{figure}[H]
\caption{The differences in standard errors of the estimated impulse responses
for $h=1,...,24$ steps ahead for four LP models and the AR(1)-GARCH(1,1), with
 $T=500,1000,2000,5000$, $\beta_{1}=0.9$, $\alpha_1=0.5$ and $ \alpha_2=0.3$.\label{fig:diffstderror09_alpha03}}

\centering{}\includegraphics[width=15cm,height=8cm]{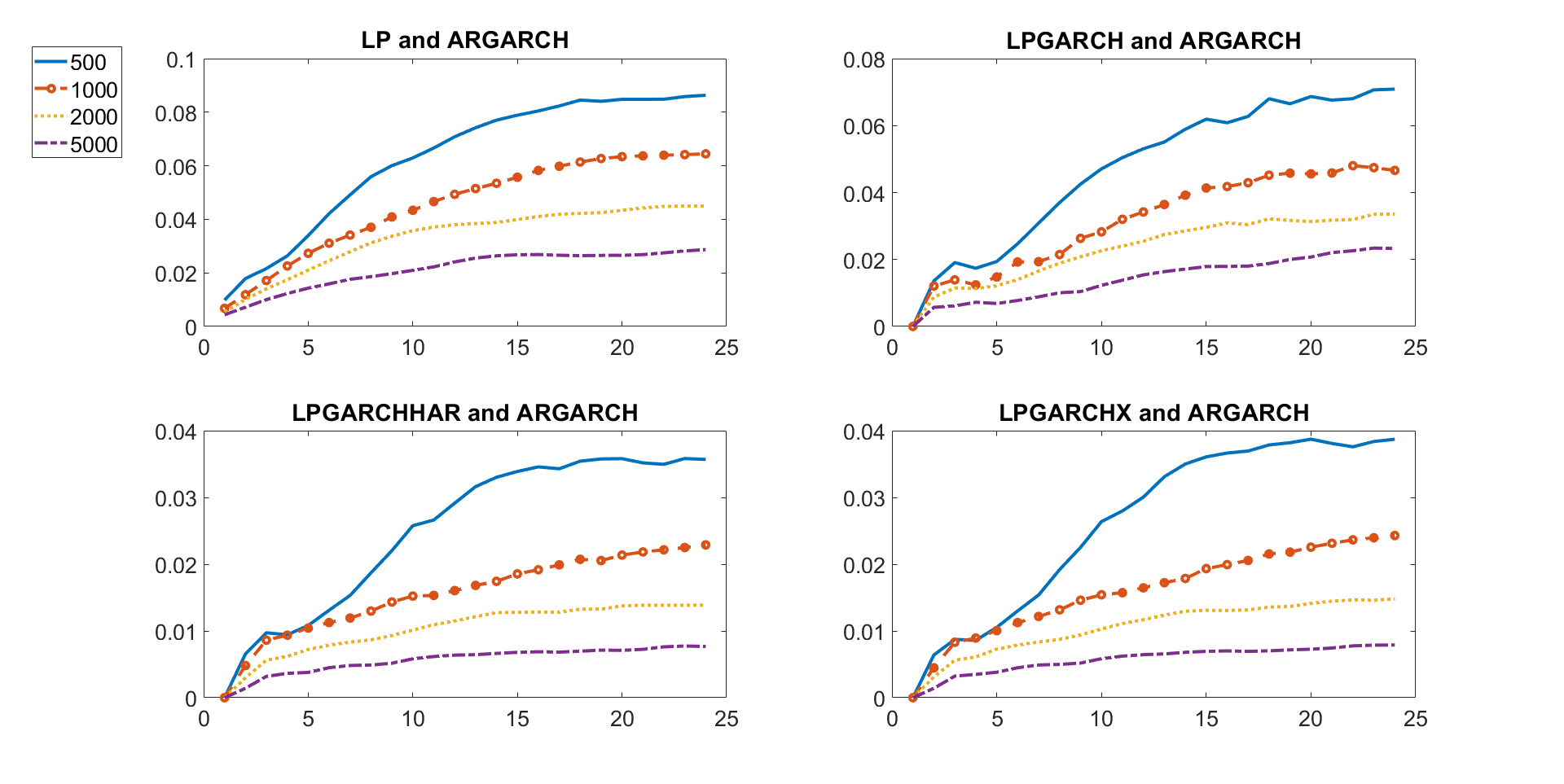}
\end{figure}

\begin{figure}[H]
\caption{The differences in standard errors of the estimated impulse responses
for $h=1,...,24$ steps ahead for four LP models and the AR(1)-GARCH(1,1), with
 $T=500,1000,2000,5000$, $\beta_{1}=0.95$, $\alpha_1=0.5$ and $ \alpha_2=0.3$.\label{fig:diffstderror095_alpha03}}

\centering{}\includegraphics[width=15cm,height=8cm]{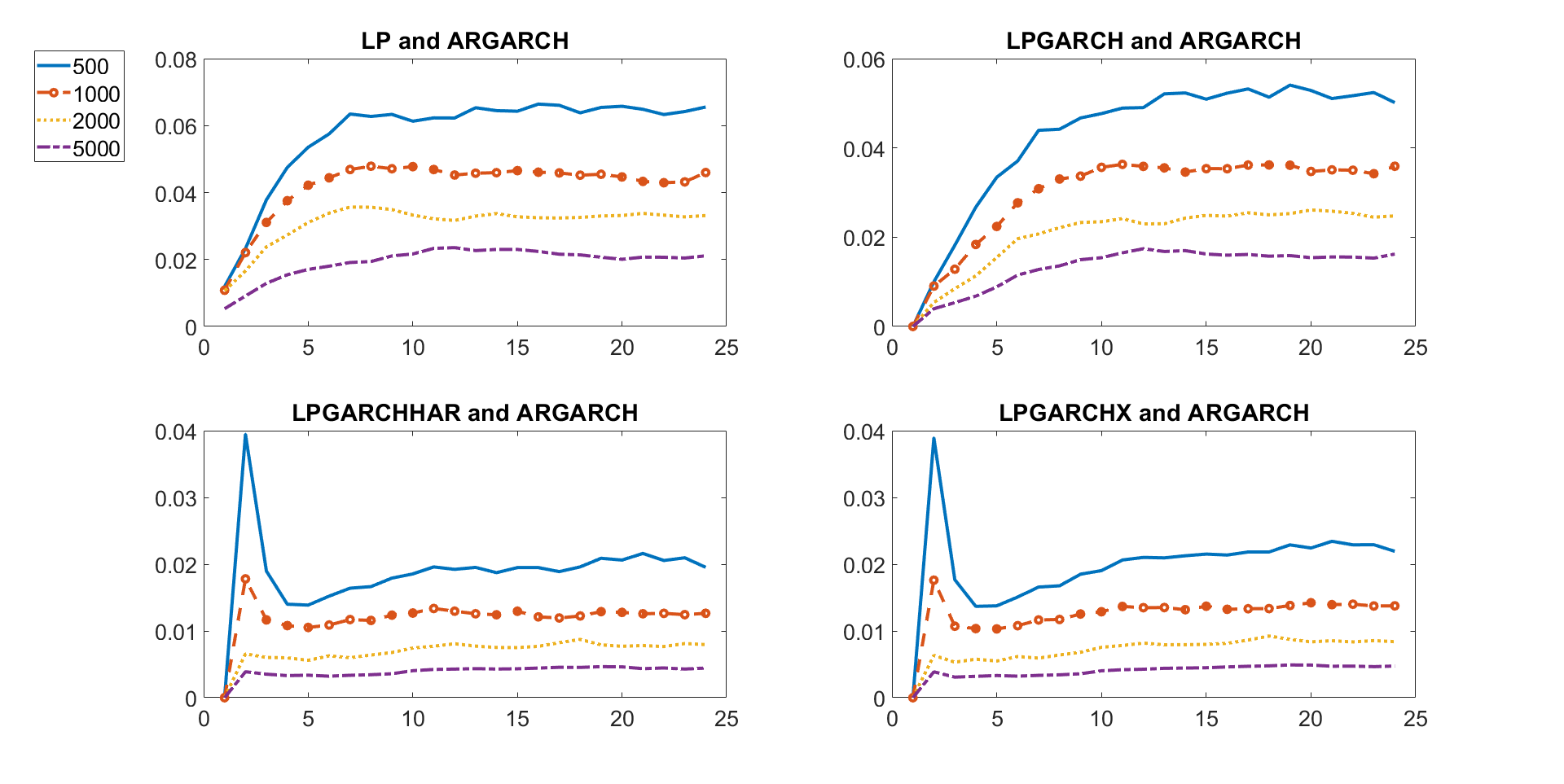}
\end{figure}

\setcounter{figure}{0}

\section{Additional Figures and Tables for $\alpha_2=0.48$\label{sec:Additionalresultsalpha2_0.48}}

\begin{table}[H]
\caption{Mean standard errors of various Local Projection models relative to the AR(1)-GARCH(1,1), averaged over a 24-step forecast horizon. The GARCH data-generating process (DGP) is defined by $\gamma=1$, $\alpha_1 = 0.5 $, and $\alpha_2=0.48$.\label{tab:Root-Mean-Square LP (alpha2_048)}}

\centering{}%
\begin{tabular}{cccccc}
\hline 
$\beta_{1}$ & T & LP & LP-GARCH & LP-GARCH-HAR & LP-GARCHX\tabularnewline
\hline 
0.6 & 500 & $0.0881$ & $0.0494$ & $0.0289$ & $0.0291 $\tabularnewline
 & 1000 & $0.0683$ & $0.0360$ & $0.0205$ & $0.0213$\tabularnewline
 & 2000 & $0.0607$ & $0.0264$ & $0.0136$ & $0.0140$\tabularnewline
 & 5000 & $0.0655$ & $0.0177$ & $0.0093$ & $0.0097$\tabularnewline
\hline 
0.8 & 500 & $0.0759$ & $0.0404$ & $0.0262$ & $0.0269 $\tabularnewline
 & 1000 & $0.0715$ & $0.0283$ & $0.0186$ & $0.0193$\tabularnewline
 & 2000 & $0.0699$ & $0.0217$ & $0.0129$ & $0.0136$\tabularnewline
 & 5000 & $0.0642$ & $0.0133$ & $0.0075$ & $0.0082$\tabularnewline
\hline 
0.9 & 500 & $0.0878$ & $0.0396$ & $0.0239$ & $0.0252 $\tabularnewline
 & 1000 & $0.0761$ & $0.0268$ & $0.0150$ & $0.0163$\tabularnewline
 & 2000 & $0.0681$ & $0.0190$ & $0.0105$ & $0.0111$\tabularnewline
 & 5000 & $0.0633$ & $0.0117$ & $0.0069$ & $0.0075$\tabularnewline
\hline 
0.95 & 500 & $0.0878$ & $0.0351$ & $0.0194$ & $0.0209 $\tabularnewline
 & 1000 & $0.0648$ & $0.0240$ & $0.0153$ & $0.0157$\tabularnewline
 & 2000 & $0.0628$ & $0.0164$ & $0.0076$ & $0.0084$\tabularnewline
 & 5000 & $0.0511$ & $0.0104$ & $0.0052$ & $0.0055$\tabularnewline
\hline 
\end{tabular}
\end{table}

\begin{figure}[H]
\caption{The standard errors of the estimated impulse responses for $h=1,...,24$
steps ahead for four LP models and the AR(1)-GARCH(1,1), with
$T=500,1000,2000,5000$, $\beta_{1}=0.6$, $\alpha_1=0.5$ and $ \alpha_2=0.48$.\label{fig:Thestderrors06_alpha048}}
\centering{}\includegraphics[width=15cm,height=8cm]{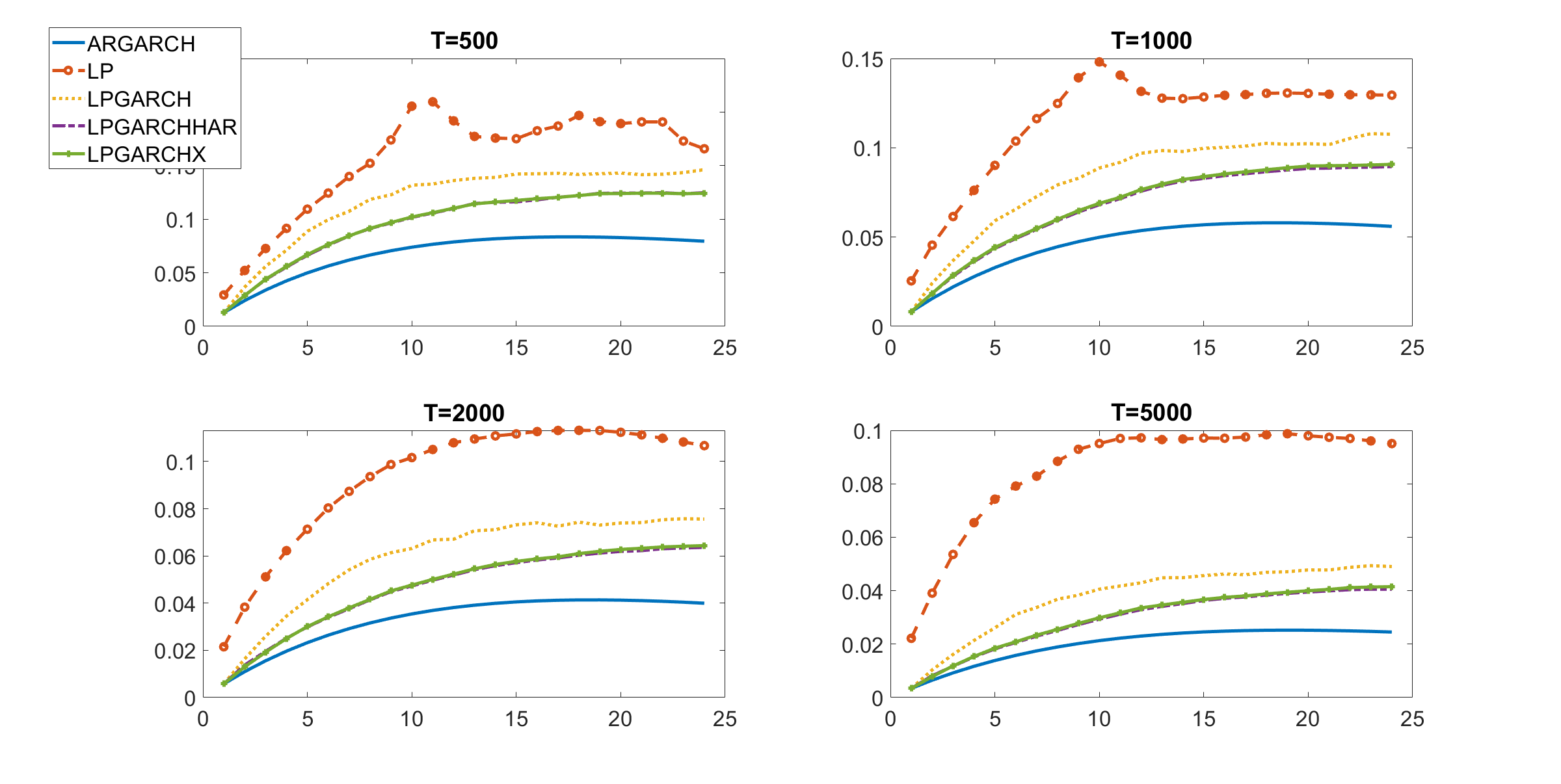}
\end{figure}

\begin{figure}[H]
\caption{The standard errors of the estimated impulse responses for $h=1,...,24$
steps ahead for four LP models and the AR(1)-GARCH(1,1), with
$T=500,1000,2000,5000$, $\beta_{1}=0.8$, $\alpha_1=0.5$ and $ \alpha_2=0.48$.\label{fig:Thestderrors08_alpha048}}

\centering{}\includegraphics[width=15cm,height=8cm]{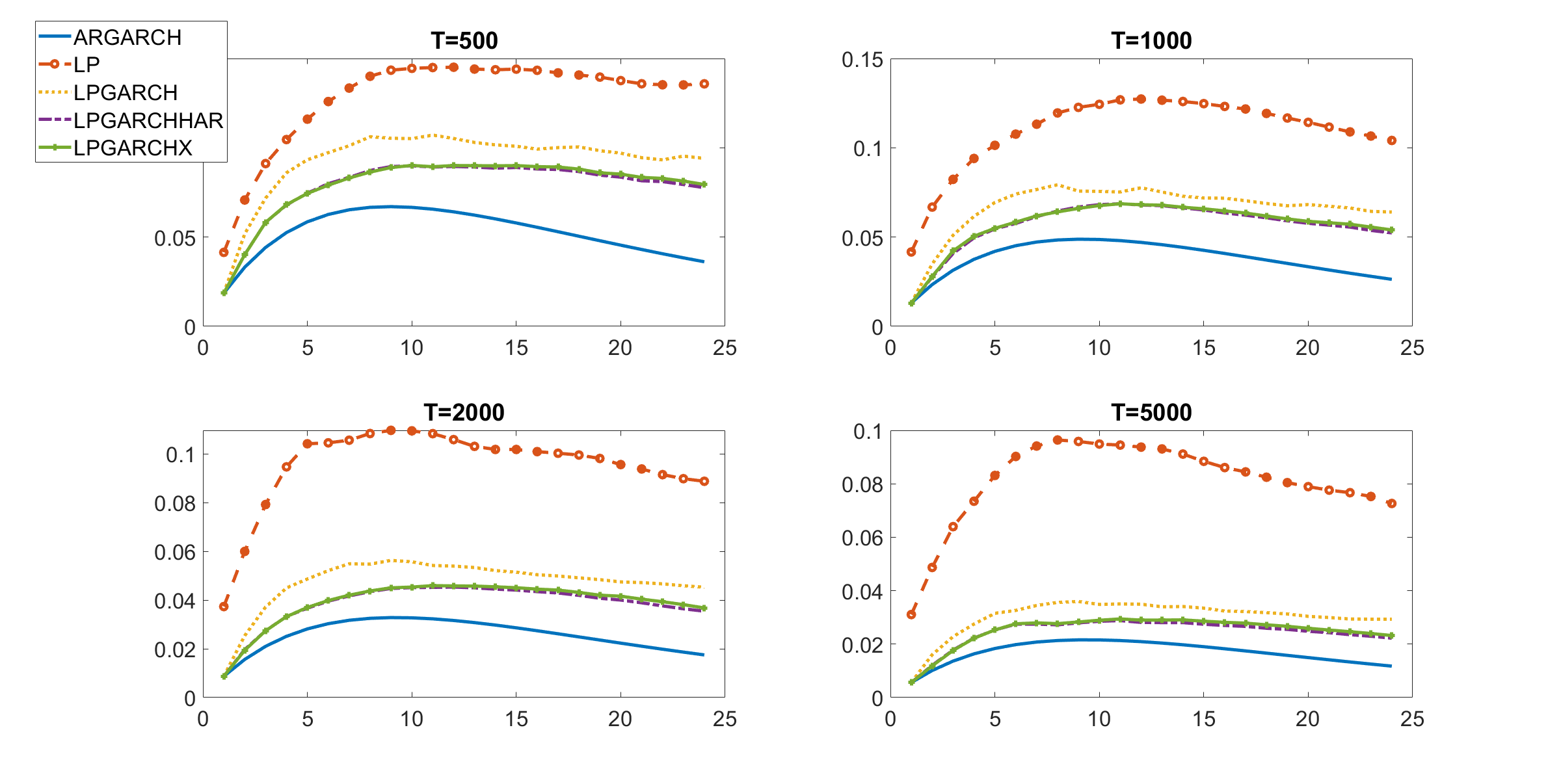}
\end{figure}

\begin{figure}[H]
\caption{The standard errors of the estimated impulse responses for $h=1,...,24$
steps ahead for four LP models and the AR(1)-GARCH(1,1), with
$T=500,1000,2000,5000$, $\beta_{1}=0.9$, $\alpha_1=0.5$ and $ \alpha_2=0.48$.\label{fig:Thestderrors09_alpha048}}

\centering{}\includegraphics[width=15cm,height=8cm]{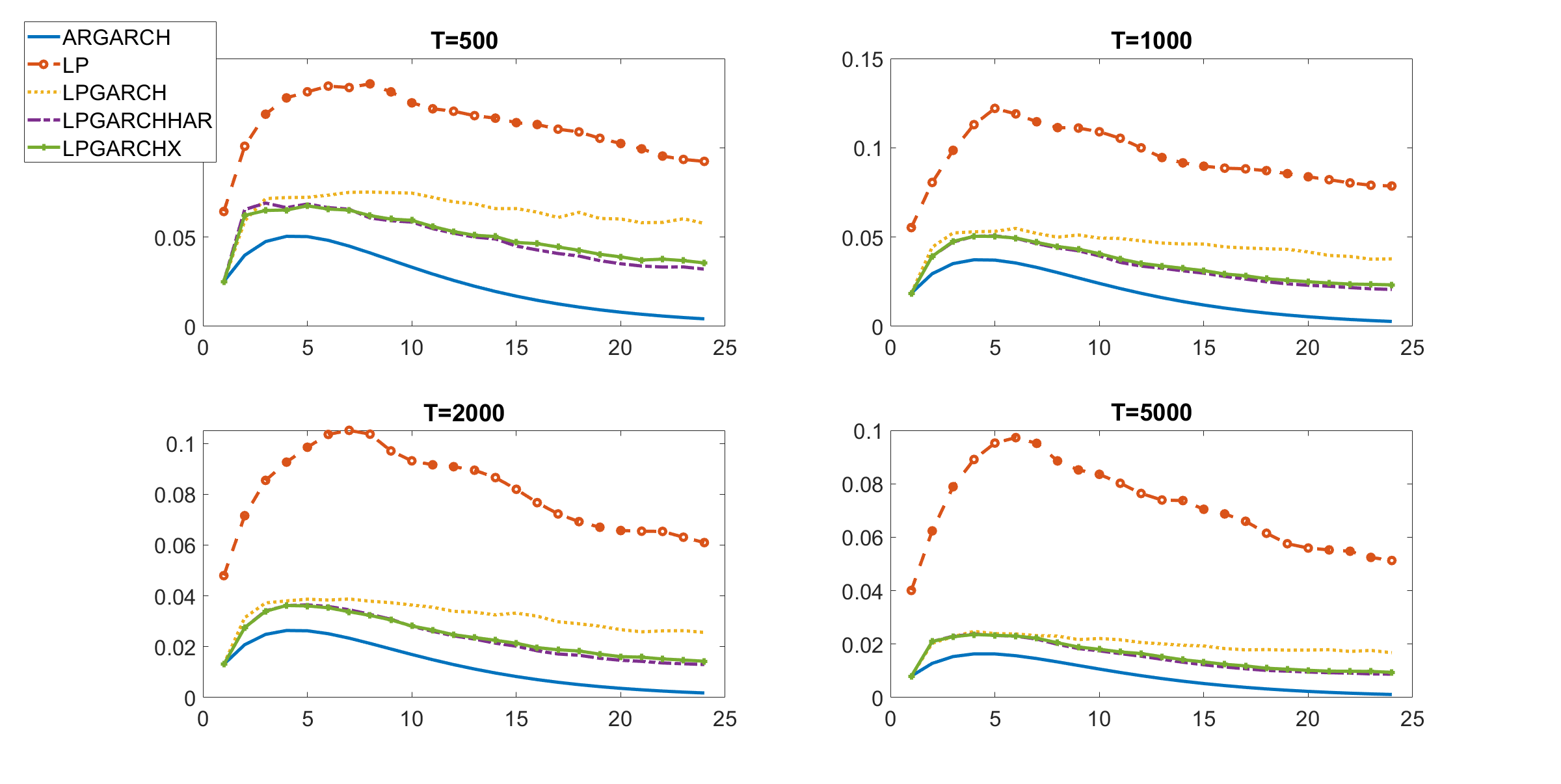}
\end{figure}

\begin{figure}[H]
\caption{The standard errors of the estimated impulse responses for $h=1,...,24$
steps ahead for four LP models and the AR(1)-GARCH(1,1), with
$T=500,1000,2000,5000$, $\beta_{1}=0.95$, $\alpha_1=0.5$ and $ \alpha_2=0.48$.\label{fig:Thestderrors095_alpha048}}

\centering{}\includegraphics[width=15cm,height=8cm]{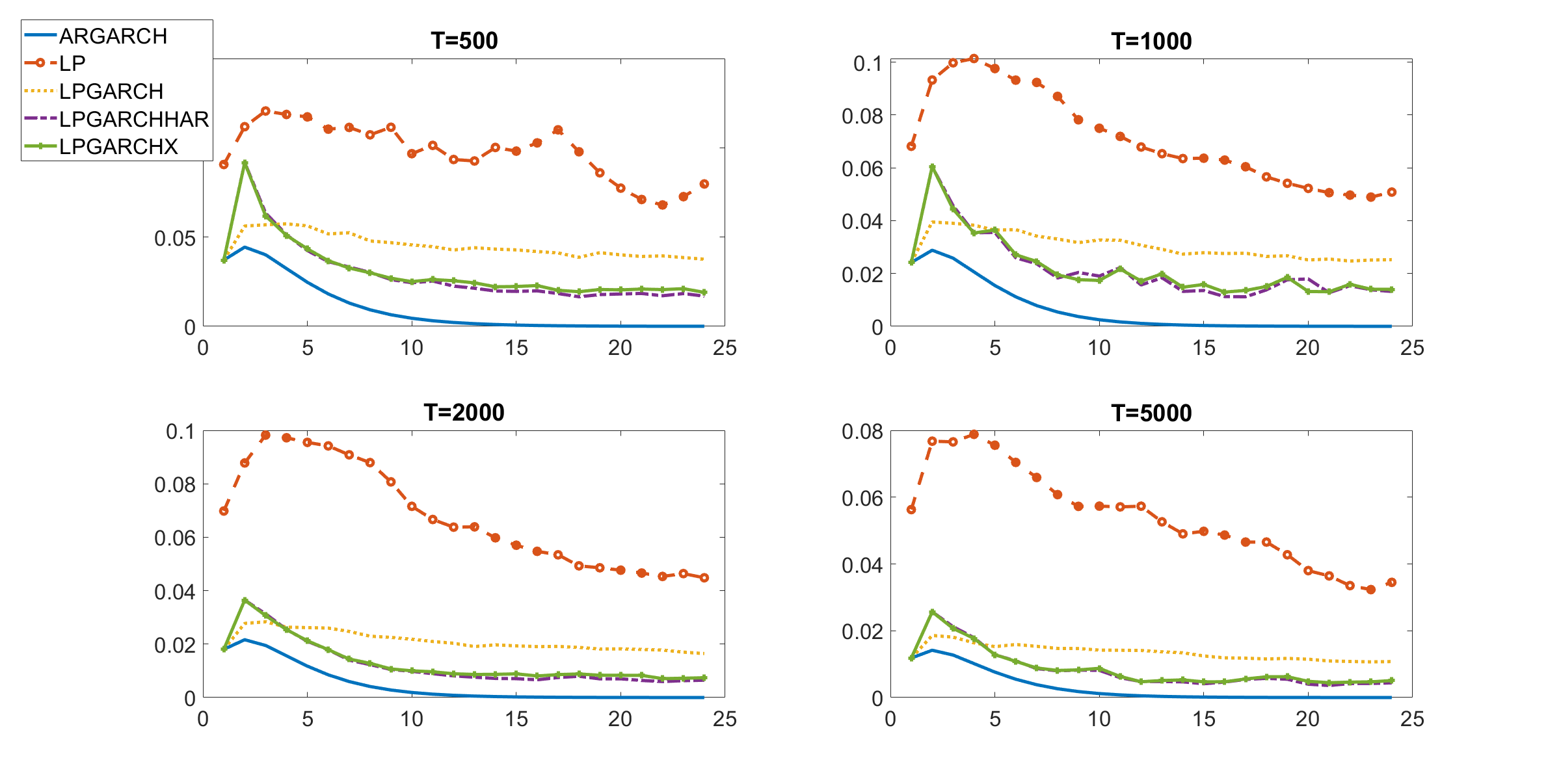}
\end{figure}

\begin{figure}[H]
\caption{The differences in standard errors of the estimated impulse responses
for $h=1,...,24$ steps ahead for four LP models and the AR(1)-GARCH(1,1), with
 $T=500,1000,2000,5000$, $\beta_{1}=0.6$, $\alpha_1=0.5$ and $ \alpha_2=0.48$.\label{fig:diffstderror06_alpha048}}

\centering{}\includegraphics[width=15cm,height=8cm]{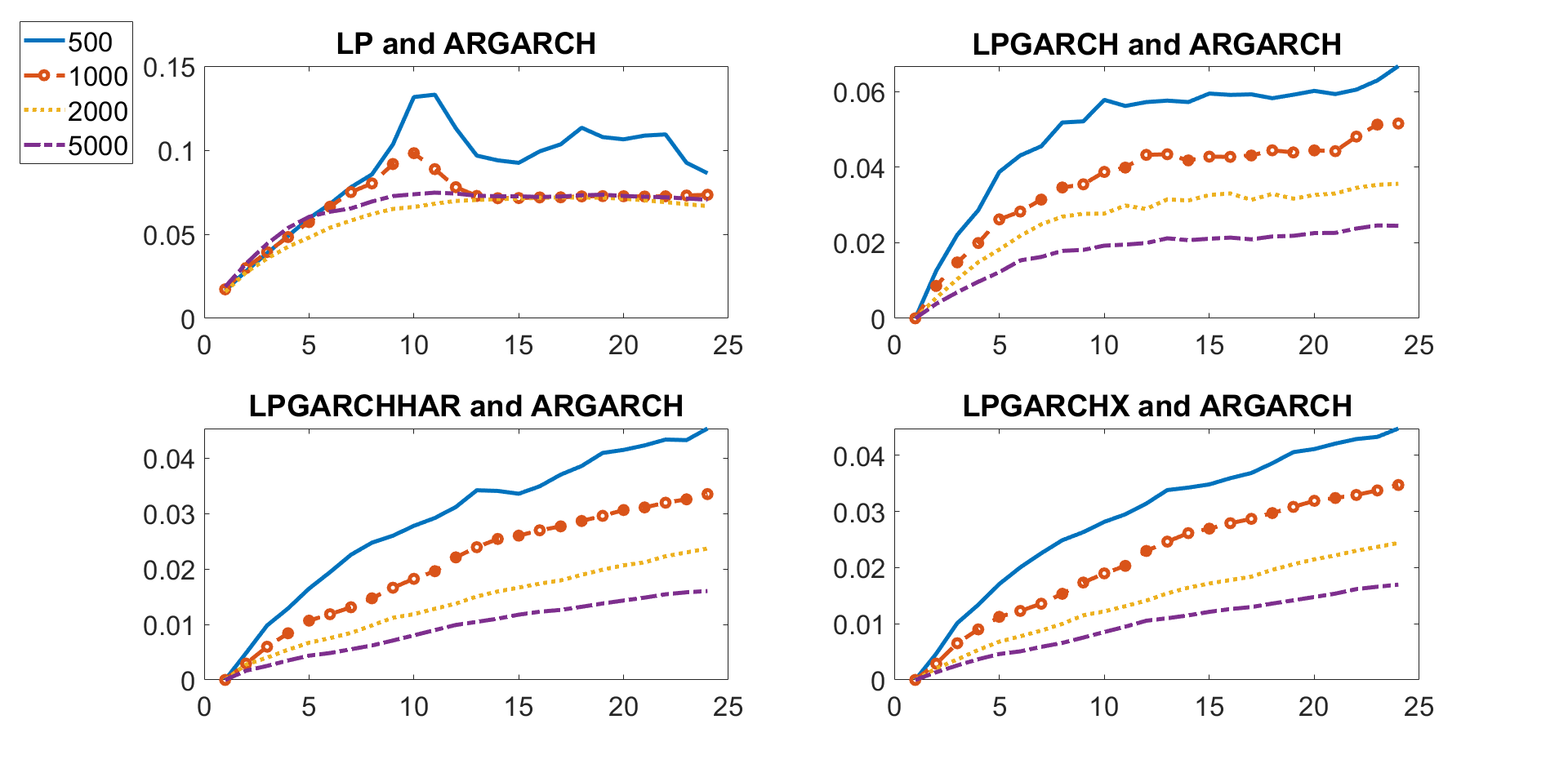}
\end{figure}

\begin{figure}[H]
\caption{The differences in standard errors of the estimated impulse responses
for $h=1,...,24$ steps ahead for four LP models and the AR(1)-GARCH(1,1), with
 $T=500,1000,2000,5000$, $\beta_{1}=0.8$, $\alpha_1=0.5$ and $ \alpha_2=0.48$.\label{fig:diffstderror08_alpha048}}

\centering{}\includegraphics[width=15cm,height=8cm]{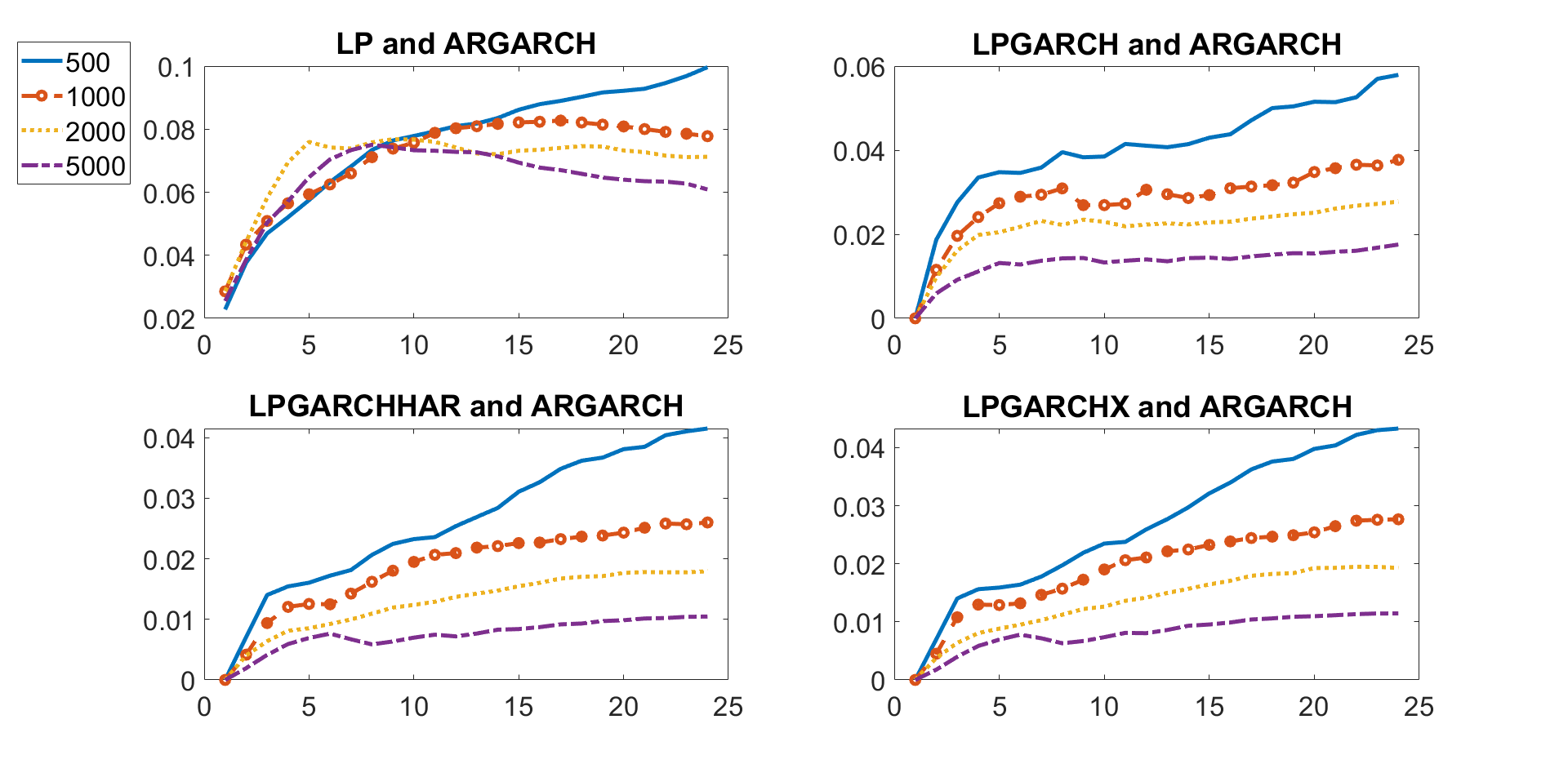}
\end{figure}

\begin{figure}[H]
\caption{The differences in standard errors of the estimated impulse responses
for $h=1,...,24$ steps ahead for four LP models and the AR(1)-GARCH(1,1), with
 $T=500,1000,2000,5000$, $\beta_{1}=0.9$, $\alpha_1=0.5$ and $ \alpha_2=0.48$.\label{fig:diffstderror09_alpha048}}

\centering{}\includegraphics[width=15cm,height=8cm]{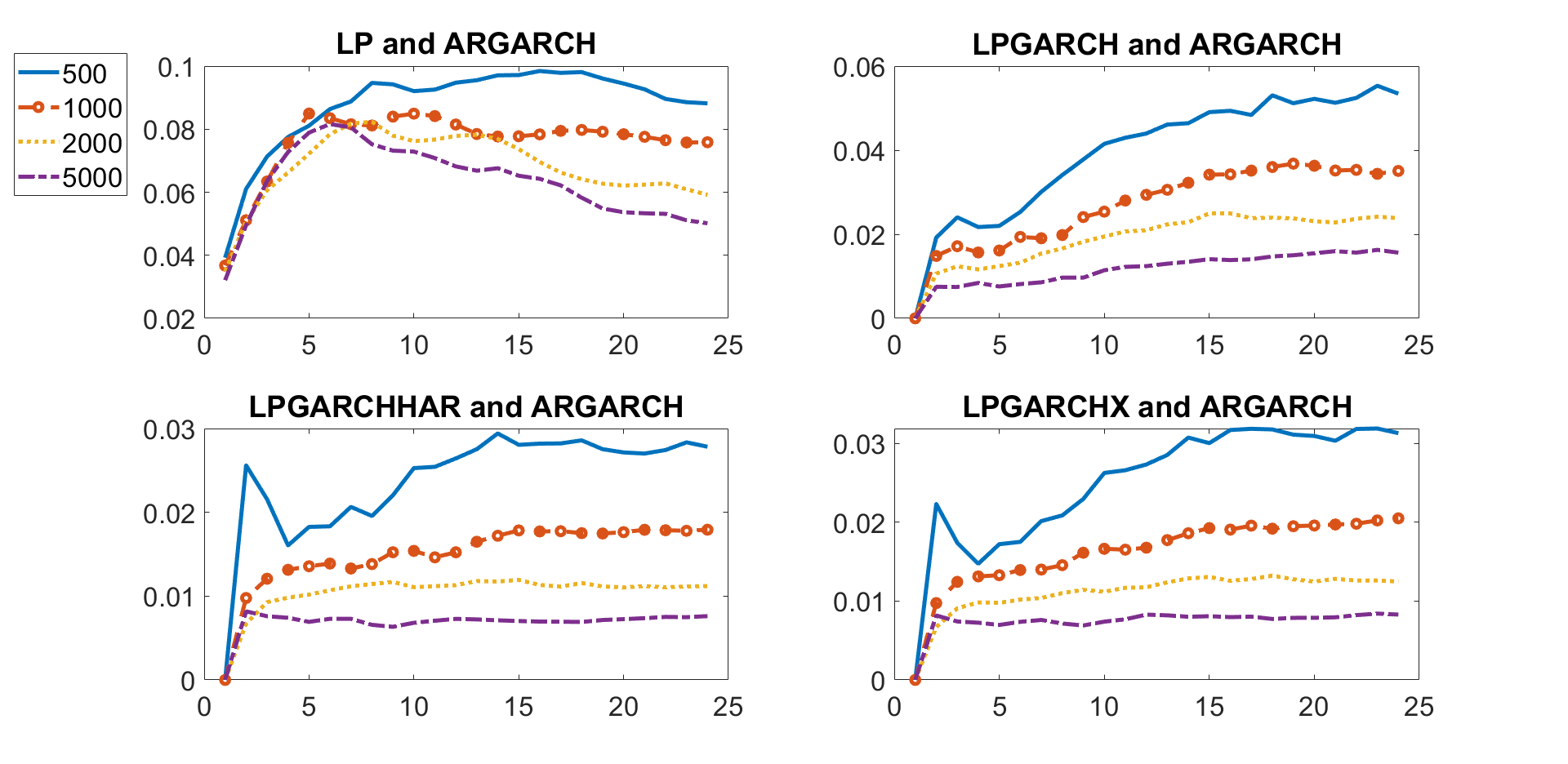}
\end{figure}

\begin{figure}[H]
\caption{The differences in standard errors of the estimated impulse responses
for $h=1,...,24$ steps ahead for four LP models and the AR(1)-GARCH(1,1), with
 $T=500,1000,2000,5000$, $\beta_{1}=0.95$, $\alpha_1=0.5$ and $ \alpha_2=0.48$.\label{fig:diffstderror095_alpha048}}

\centering{}\includegraphics[width=15cm,height=8cm]{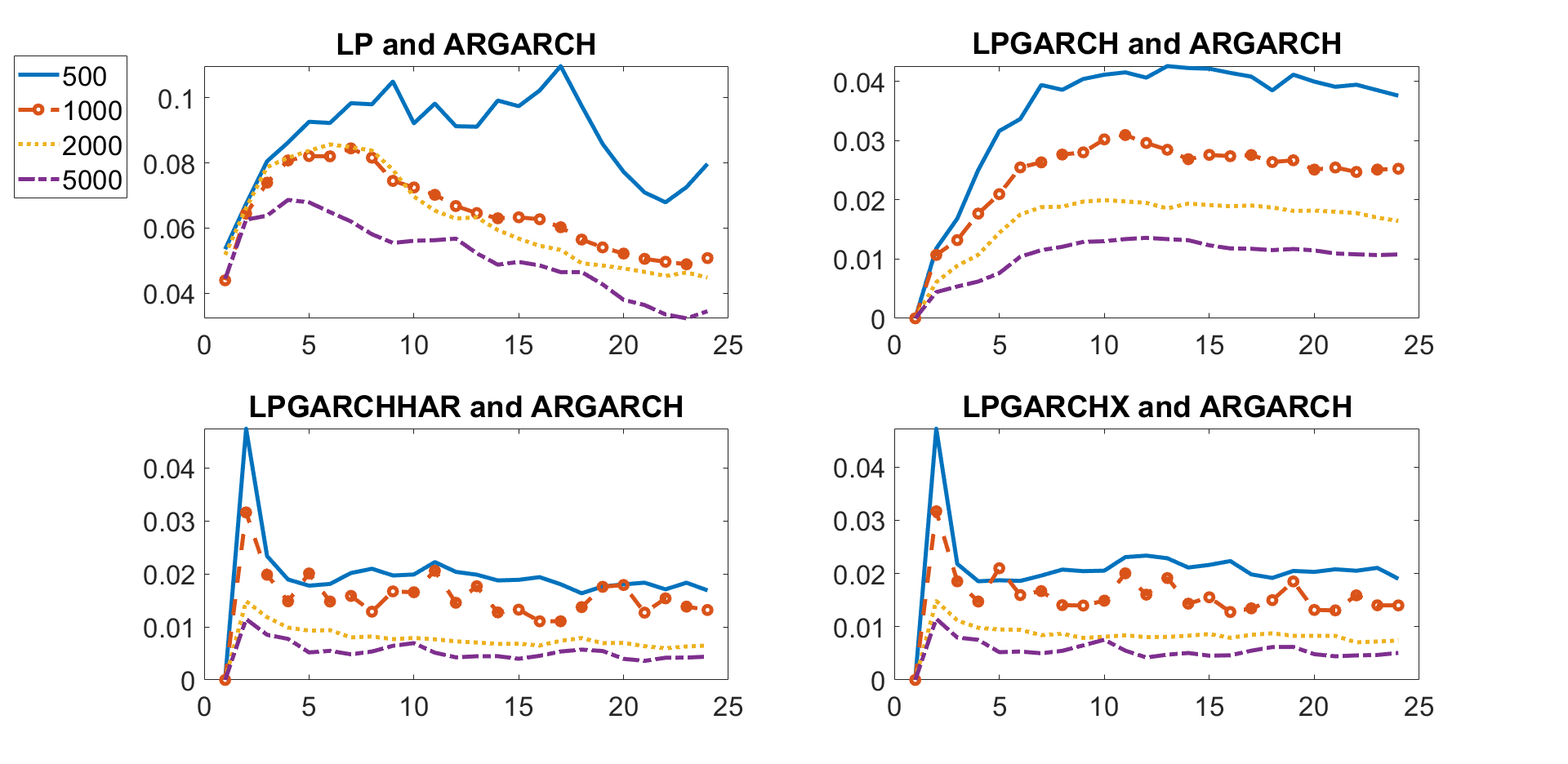}
\end{figure}

\bibliography{LP_GARCH_BiB}

\end{document}